\documentclass[%
 aip,
 amsmath,amssymb,
 reprint,%
]{revtex4-1}

\usepackage{graphicx}
\usepackage{dcolumn}
\usepackage{bm}
\usepackage[utf8]{inputenc}
\usepackage[T1]{fontenc}
\usepackage{mathptmx}
\usepackage{etoolbox}
\usepackage{color}
\usepackage[normalem]{ulem}
\usepackage{hyperref}

\usepackage{bm}
\DeclareSymbolFont{cmsymbols}{OMS}{cmsy}{m}{n}
\SetSymbolFont{cmsymbols}{bold}{OMS}{cmsy}{b}{n}
\DeclareSymbolFontAlphabet{\mathcal}{cmsymbols}

\definecolor{si}{rgb}{0,0,0}
\definecolor{review}{rgb}{0,0,0}
\newcommand{\SI}[1]{{\color{si} #1}}
\newcommand{\rev}[1]{{\color{review} #1}}

\makeatletter
\def\@email#1#2{%
 \endgroup
 \patchcmd{\titleblock@produce}
  {\frontmatter@RRAPformat}
  {\frontmatter@RRAPformat{\produce@RRAP{*#1\href{mailto:#2}{#2}}}\frontmatter@RRAPformat}
  {}{}
}%
\makeatother
\begin{document}

\preprint{AIP/123-QED}

\title{A Variational Approach to Assess Reaction Coordinates for Two-Step Crystallisation}
\author{A. R. Finney}
 \email{a.finney@ucl.ac.uk; m.salvalaglio@ucl.ac.uk}
 \affiliation{Thomas Young Centre and Department of Chemical Engineering, University College London, London WC1E 7JE, United Kingdom}
\author{M. Salvalaglio}%
\affiliation{Thomas Young Centre and Department of Chemical Engineering, University College London, London WC1E 7JE, United Kingdom}%


\date{\today}

\begin{abstract}
Molecule- and particle-based simulations provide the tools to test, in microscopic detail, the validity of \emph{classical nucleation theory}. 
In this endeavour, determining nucleation mechanisms and rates for phase separation requires an appropriately defined reaction coordinate to describe the transformation of an out-of-equilibrium parent phase, for which myriad options are available to the simulator.
In this article, we describe the application of the variational approach to Markov processes (VAMP) to quantify the suitability of reaction coordinates to study crystallisation from supersaturated colloid suspensions.
Our analysis indicates that collective variables (CVs) that correlate with the number of particles in the condensed phase, the system potential energy and approximate configurational entropy often feature as the most appropriate order parameters to quantitatively describe the crystallisation process.
We apply time-lagged independent component analysis to reduce high-dimensional reaction coordinates constructed from these CVs to build Markov State Models (MSMs), which indicate that two barriers separate a supersaturated fluid phase from crystals in the simulated environment.
The MSMs provide consistent estimates for crystal nucleation rates, regardless of the dimensionality of the order parameter space adopted; however, the two-step mechanism is only consistently evident from spectral clustering of the MSMs in higher dimensions.
As the method is general and easily transferable, the variational approach we adopt could provide a useful framework to study controls for crystal nucleation.

\end{abstract}

\maketitle

\section{\label{sec:introduction}Introduction}
Crystal nucleation marks the emergence of long-range order in a parent liquid or gas phase which may only display short-range symmetry at the scale of constituent monomers.
In particle or molecular systems, the size of the critical nucleus---the smallest collection of monomers with crystalline order that can lead to bulk crystals---is typically many orders of magnitude smaller than Avogadro's number. \cite{Mullin2001Crystallization}
Combined with the fact that nucleation is a rare event, this makes investigating nucleation mechanisms \emph{in situ} particularly challenging.

Computer simulations employing Molecular Dynamics (MD) algorithms have provided significant insights into crystallisation pathways, especially since the advent of methods to enhance the sampling of rare events. \cite{Sosso2016CrystalSimulations}
To monitor the crystallisation process and establish nucleation kinetics in these types of simulations, a suitable reaction coordinate (RC) is needed to reduce the $6N$ ($N$ being the number of monomers) dimensional phase space to just a handful of collective variables (CVs) that completely capture the emergence of long-range order. \cite{Peters2016ReactionTests,Blow2021TheDetails}
All other degrees of freedom can be ignored when determining \emph{relative} nucleation rates.

Classical Nucleation Theory (CNT) adopts the size of an embryo of a new thermodynamic phase, usually its radius, as an RC for phase transformation.\cite{Kashchiev2000Nucleation:Applications}
With respect to crystallisation, the number of monomers in the new phase is a more appropriate metric for this size, given the highly faceted and non-spherical geometry of crystals, even at small sizes.
As several studies have demonstrated, however, a one-dimensional RC can be unsuitable to describe the evolution of a crystallising system. \cite{Wolde1997EnhancementFluctuations,Jiang2019NucleationSpinodal,Kashchiev2020ClassicalCrystals,Bulutoglu2022AnClusters}
For example, our own work demonstrates that a two-dimensional RC, quantifying both the size of emerging clusters and their crystalline order, is helpful to describe the formation of crystals from metastable solutions.\cite{Salvalaglio2015UreaDependent,Salvalaglio2015Molecular-dynamicsSolution,Finney2022MultipleNucleation}
Still, no obvious definition for these variables can have consequences for understanding nucleation mechanisms and predicting crystallisation rates.\cite{Peters2016ReactionTests,Blow2021TheDetails,Zimmermann2018NaClEstimates}

Common CVs to approximate RCs for crystallisation are functions of the positional coordinates of a collection of particles.
These must be continuous and differentiable if used in biased enhanced sampling schemes, but typically this is not a prerequisite for analysis purposes.
A simple example CV used in this context is the first-sphere coordination number; however, this typically fails to capture the local symmetry of a crystal lattice and, therefore, might not be suitable to distinguish dense amorphous phases and crystal polymorphs.
Bond orientational order parameters can achieve this by, for example, making use of spherical harmonic functions to quantify the relative position of monomers in a coordination sphere with respect to one another.\cite{Steinhardt1983Bond-orientationalGlasses}
Alternatively, if a reference structure is known, one can compute the relative distance between particles in simulations and this reference in topography space, or perform topological graph analyses, with nodes in the graph representing monomers, to identify crystal structures.\cite{Larsen2016RobustMatching,Francia2020SystematicLandscapes}
Accurate classification of monomers at crystal surfaces and defects is challenging in all of these methods due to under-coordination at these sites.

No generally applicable procedure exists to choose order parameters to study multi-step crystal nucleation; this often comes down to chemical/physical intuition on behalf of the researcher.
A useful review on the topic was provided by Peters,\cite{Peters2016ReactionTests} who remarks ``[h]uman intuition
remains the best source of trial coordinates and mechanistic hypotheses, and there is no procedure
for having an epiphany.''
There are, however, methods available to test the suitability of the RC. 
These include, for example, likelihood maximisation\cite{Peters2006ObtainingMaximization} and committor analyses \cite{Geissler1999KineticWater}.

The Variational Approach to Markov Processes (VAMP)\cite{Wu2020VariationalData}is a generalised version of the Variational Approach to Conformational Dynamics (VAC)\cite{Noe2013ASystems} that has been successfully applied to determine suitable RCs in systems with stochastic dynamics, including protein folding and problems associated with molecular and crystallisation kinetics. \cite{Zhang2019ImprovingCrystallization,Mardt2018VAMPnetsKinetics}
Here, we apply VAMP to test the suitability of thousands of potential RCs defined by combining sets of CVs typically used to study crystallisation pathways in monoatomic solids. 
VAMP allows us to quantify the effectiveness of the RCs to capture the slow dynamic modes associated with crystallisation and identify which combinations of parameters best describe emerging order.
To this aim, we perform simulations of metastable colloid suspensions which undergo crystallisation, use VAMP to identify the most suitable combination of CVs for every set of dimensionality to define RCs, perform dimensionality reduction using TICA, and construct Markov State Models to quantify kinetics and transformation mechanisms. 
In the following section, we provide a brief overview of the salient features of the methods employed, with an emphasis on VAMP. For a more involved discussion, including associated Markov modelling methods, see References \citenum{Wu2020VariationalData}, \citenum{2013AnSimulation} and \citenum{Noe2015KineticSimulation}.

\section{\label{sec:theory}Theoretical Background}

Projections of the highly nonlinear evolution of a system in phase space onto a low-dimensional representation are often employed to understand physicochemical processes.
When analysing transitions in nonlinear dynamical systems, the Koopman operator, $\mathcal{K}$, is linear in a space of infinite observables, which completely describes the time evolution of a system.
If a system occupies states in phase space at $\mathbf{x}_1$ at time $t$, $\mathcal{K}$ is an operator which acts on the function $g$ to determine the expectation of the system being in states at $\mathbf{x}_2$ at $t+\tau$ ($\tau$ being some lag-time), given the conditional probability density of states, $p(\mathbf{x}_1,\mathbf{x}_2)$:
\begin{equation}
    [\mathcal{K}g](\mathbf{x}) = \int p(\mathbf{x}_1,\mathbf{x}_2) g(\mathbf{x}_2) \; d\mathrm{x}_2 = \mathbb{E}[g(\mathbf{x}_{t+ \tau})]
\end{equation}
Its spectral decomposition, therefore, completely characterises (meta)stable states and transitions between them.\cite{Noe2013ASystems}
With a finite number of functions characterising the time evolution for the process of interest, a good approximation of $\mathcal{K}$ is the propagator, $\mathbf{K}$, which in principle allows determination of the transition probabilities and timescales associated with crystallisation in closed thermodynamic systems (where the partition function is bounded by the finite number of particles in the simulations).

Given a set of functions of the configurational space of the system of interest, $\mathbf{f}(\mathbf{r})=(f_1(\mathbf{r}), f_2(\mathbf{r}),\dots,f_n(\mathbf{r}))$, i.e., CVs which project the full $3N$ configurational coordinates of $N$ atoms/particles in a system, $\mathbf{r}$, onto an $n$-dimensional RC, the time-dependent Markovian dynamics can be predicted according to,
\begin{equation}
    \mathbb{E} [\mathbf{f}(\mathbf{r},t+ \tau) ] \approx \mathbf{K} ^\top \mathbb{E} [\mathbf{f}(\mathbf{r},t) ]
    \label{eqn:markov}
\end{equation}
where $\mathbb{E}$ is the expectation value evaluated for an average trajectory, and $\mathbf{f}$ is the array of CVs that approximates the eigenfunctions characterising transitions between (meta)stable states.
VAMP can be applied to optimise the dimensionality and choice of $\mathbf{f}$.
This involves computing the time-dependent covariance matrices,
\begin{equation}
\begin{split}
    \mathbf{C}_{00} &= \frac{1}{T-\tau} \sum_{t=0}^{T-\tau} [\mathbf{f}(\mathbf{r},t) - \overline{\mathbf{f}_{0}}(\mathbf{r})][\mathbf{f}(\mathbf{r},t) - \overline{\mathbf{f}_{0}}(\mathbf{r})] \\
    \mathbf{C}_{11} &= \frac{1}{T-\tau} \sum_{t=\tau}^{T} [\mathbf{f}(\mathbf{r},t) - \overline{\mathbf{f}_{1}}(\mathbf{r})][\mathbf{f}(\mathbf{r},t) - \overline{\mathbf{f}_{1}}(\mathbf{r})] \\
    \mathbf{C}_{01} &= \frac{1}{T-\tau} \sum_{t=0}^{T-\tau} [\mathbf{f}(\mathbf{r},t) - \overline{\mathbf{f}_{0}}(\mathbf{r})][\mathbf{f}(\mathbf{r},t+\tau) - \overline{\mathbf{f}_{1}}(\mathbf{r})]    
\end{split}
\end{equation}
where $\overline{\mathbf{f}_{0}}(\mathbf{r})$ and $\overline{\mathbf{f}_{1}}(\mathbf{r})$ are the mean CV values in $t=0 \rightarrow (T- \tau)$ and $t=\tau \rightarrow T$ time windows, respecively.
$\mathbf{K}$ is simply given by $\mathbf{C}_{00}^{-1}\mathbf{C}_{01}$, and singular values of the half-weighted propagator provide a metric to determine how well $\mathbf{f}(\mathbf{r})$ approximates the eigenvectors that capture the slow modes for crystallisation:
\begin{equation}    
    R_2 (\mathbf{f},\mathbf{r}) = || \mathbf{C}_{00}^{-\frac{1}{2}} \, \mathbf{C}_{01} \, \mathbf{C}_{11} ^{-\frac{1}{2}} ||^2
\end{equation}
$R_2$ is known as the VAMP-2 score and is used in this work to judge the appropriateness of RCs to describe two-step colloid crystallisation.
Cross-validation\cite{McGibbon2015VariationalKinetics} is carried out to ensure that the dynamical model is not overfitted. This is done by partitioning the CV time series into training and test data, building the model on the training subset and validating it with the remaining data. 
It is then possible to compute $R_2$ from multiple partitions of training and test data. \cite{Wu2020VariationalData}

In the following sections, we use VAMP to identify the best combinations of CVs to determine the crystallisation kinetics.
In stochastic dynamical systems, one can construct models based on Markovian dynamics to map the evolution of systems in these low-dimensional coordinates to extract mechanistic insight and timescales for the processes of interest.
First, we use time-lagged independent component analysis (TICA)\cite{Schwantes2013ImprovementsNTL9,Perez-Hernandez2013IdentificationConstruction} to project the CVs onto one or two components by solving,
\begin{equation}
    \mathbf{C}(\tau) \mathbf{f}(\mathbf{r},t) = \lambda\mathbf{C}_{00}\mathbf{f}(\mathbf{r},t)
\end{equation}
where $\lambda$ are  eigenvalues which determine the slowness of the transitions in the system dynamics.

After TICA, we partition the sampled data into discrete partitions \{$s_2, \; s_2, \dots, s_N$\}.
The probability weights for each partition are a function of the stationary distribution, $\mu$:
\begin{equation}
    \pi_i = \int_{\mathbf{x} \in s_i} \mu( \mathbf{x}) \; d\mathbf{x}
\end{equation}
so that combining simulation trajectories that sample reasonably well the local TICA space centred on different partitions allows us to determine the relative probability for the system to occupy these partitions and evaluate free energy differences.
In addition, counting the transitions between partitions in the TICA trajectories allows us to evaluate $\mathbf{K}$:
\begin{equation}
    \mathbf{K}(\tau) \mathbf{\pi} = \mathbf{\pi}
\end{equation}
and, therefore, provides kinetic information for the crystallising system.

\section{\label{sec:methods}Computational Details}
We performed simulations using the LAMMPS (v. 7Aug2019) MD simulator\cite{Thompson2022LAMMPSScales}. 
\rev{To prepare the initial configurations,} 388 spherical particles were randomly assigned to \rev{a $10\times10\times10$} face centred cubic lattice \rev{(where the reduced lattice density was $0.005 \sigma^{-3}$ and the lattice constant was $9.283177 \sigma$)} in a simulation cell with a reduced \rev{particle} density, $\rho^* = 0.000485$.
\rev{The resulting cubic simulation cell lengths were $92.83177 \sigma$ and $\sim90$\% of the} lattice sites were vacant.
Particle interactions were modelled using a colloid/Yukawa\cite{Everaers2003InteractionEllipsoids,Safran2018StatisticalMembranes} potential to simulate van der Walls attraction and electrostatic repulsion between colloid particles \rev{in simulations adopting three-dimensional periodic boundaries}. 
The pair potential coefficients with the force field implemented in LAMMPS were $A^*=53$, $d^*=5$, $B^*=20$, representing the Hamaker constant, particle diameter, and prefactor of the Yukawa potential, which approximates a surrounding electrolyte solution as a continuum field.
Interactions were truncated at a reduced distance of $12.5\sigma$.
Particle velocities were assigned at random from a Maxwell-Boltzmann distribution with mean reduced temperature, $T^*=2$.
The simulations were performed for $2 \times 10^7$ steps with a timestep $\Delta t^* = 0.005$, during which particle velocity rescaling was carried out every 100 steps to maintain a constant temperature, and particle positions were recorded every 100 steps for subsequent analyses.
We performed 1,000 simulations where the initial velocity assignment was randomised, but all other simulation details remained the same.
While condensation was observed in all simulations, only 11 of the simulations resulted in crystallisation, as indicated by a potential energy per particle threshold: $E^* = -10 \epsilon$.
It was these crystallising trajectories that were used for analyses of RCs.

A total of 19 CVs were computed either during time integration or by post-processing simulation trajectories using the PLUMED software (\rev{v. 2.5.1})\cite{Tribello2014PLUMEDBird}.
These are useful indicators for phase separation and/or crystallisation and are potential order parameters that can be used to construct RCs.
Table~\ref{tab:table1} provides the list of CVs and their labels adopted herein.
The CVs can be classified into one of three categories: i) average properties of all particles in the system regardless of their local environment (ene, ent, cn.mean, Q4.mean and Q6.mean); ii) average properties of all particles in the system according to their local structure (q4.mean, laQ4.mean, q6.mean and laQ6.mean); iii) total numbers of particles according to some geometric criteria of their local structure (ncl, ncs, ncnq4, ncnq6, nclust1, non, fcc, hcp, bcc and ico).
Please see the \SI{Supplementary Materials (SM) Section S1} for a detailed description and mathematical definition of the CVs.

The VAMP and MSM analyses were performed using the PyEMMA (v. 2.5.11)\cite{Scherer2015PyEMMAModels} and deeptime (v. 0.4.1)\cite{Hoffmann2022Deeptime:Data} Python libraries.
See the Data Availability section for information on how to access interactive Python notebooks used in this work, and to download input files used to perform simulations and generate the CV time series data.

\begin{table}
\caption{\label{tab:table1}Collective variables (CVs) computed in this work}
\begin{ruledtabular}
\begin{tabular}{lr}
CV&Label\\
\hline
Mean first-sphere coordination number & cn.mean \\
Number of particles in a condensed phase (CN > 3) & ncl \\
Number of particles in a solid-like phase (CN > 6) & ncs \\
Number of particles in the largest cluster\cite{Tribello2017AnalyzingSimulations} & nclust1 \\
Mean Q4 Steinhardt bond order\cite{Steinhardt1983Bond-orientationalGlasses} & Q4.mean \\
Mean local Q4 bond order & q4.mean \\
Number of coordinated particles with local Q4 < 0.3 & ncnq4 \\
Local average\cite{Lechner2008AccurateParameters} Q4 bond order & laQ4.mean \\
Mean Q6 Steinhardt bond order\cite{Steinhardt1983Bond-orientationalGlasses} & Q6.mean \\
Mean local Q6 bond order & q6.mean \\
Number of coordinated particles with local Q6 > 0.7 & ncnq6 \\
Local average\cite{Lechner2008AccurateParameters} Q6 bond order & laQ6.mean \\
Pair entropy function\cite{Piaggi2017EnhancingSimulations} & ent \\
System potential energy & ene\\
Number of particles not identified as fcc/hcp/bcc/ico \footnote{Evaluated using polyhedral template matching (PTM).\cite{Larsen2016RobustMatching} } & non \\
Number of face-centred cubic particles $^\mathrm{a}$ & fcc \\
Number of hexagonal close-packed particles $^\mathrm{a}$ & hcp \\
Number of body-centred cubic particles $^\mathrm{a}$ & bcc \\
Number of icosahedral particles $^\mathrm{a}$ & ico \\
\end{tabular}
\end{ruledtabular}
\end{table}

\section{\label{sec:results}Results}
\subsection{\label{sec:2S}Two-step Colloid Crystallisation}
\begin{figure*}
\includegraphics[width=0.75\linewidth]{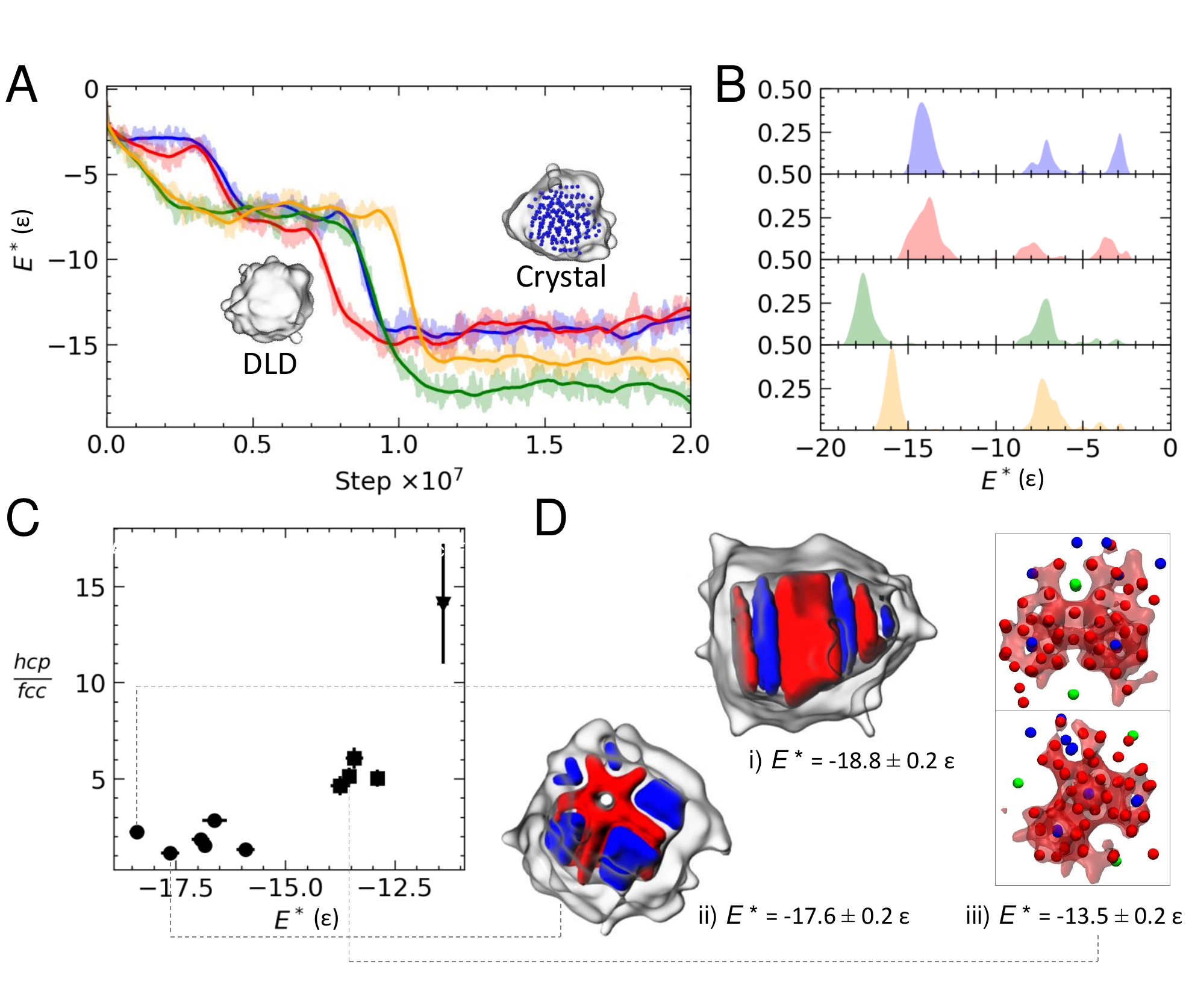}
\caption{\label{fig:ene-2step} A) Mean reduced potential energy per particle, $E^*$, as a function of step number in four different crystallising trajectories, as indicated by the colours. Raw data are shown by the shaded regions, while the solid lines result from a third-order polynomial Savitzky-Golay filtering\cite{Savitzky1964SmoothingProcedures.} of the raw values using a window length of $2 \times 10^4$ trajectory frames. Inset are snapshots of the largest cluster (indicated by the white transparent surface) in the trajectory shown by the blue curve. The blue spheres highlight colloid particles with CN > 6. B) Probability densities for $E^*$ from the trajectory data in A, as indicated by the colours. C) Average hcp/fcc fraction in crystals as a function of the average potential energy per particle in the final 5,000 frames of each trajectory. Error bars indicate uncertainties of one standard deviation in the data. Circle, square and triangle data points highlight different clusters of crystal structures. D) Snapshots of example crystal structures were taken from the end of simulation trajectories. Red domains/spheres indicate hcp-like particles, blue domains/spheres are fcc-like particles, green spheres are icosahedral-like particles, and the transparent white surface indicates all other types of particles which reside at the crystal surface. In iii, the crystal transparent surface is removed, and the fcc domain is made transparent for clarity. The same structure is projected perpendicular and parallel to the plane formed by icosahedral particles in the top and bottom boxes, respectively.}
\end{figure*}

In all 1,000 independent simulations, an initial phase separation resulted in a finite-sized droplet of a condensed disordered phase in pseudo-equilibrium with a diluted vapour-like phase characterised by a significantly lower density cf. the initial one. 
In all simulations, the emergent phase was liquid-like: colloid particles in the dense liquid droplet (DLD) were highly mobile, and there was a frequent exchange between monomers in the DLD and the surrounding low-density phases.
The condensation is indicated by a change in the average reduced potential energy per particle from an initial value of $E^* \approx -2 \epsilon$ that reaches an initial plateau corresponding to $E^* \approx -7 \epsilon$, as shown in Figure \ref{fig:ene-2step} for four example crystallising trajectories.
The time for this transition varies, as expected for an activated process of condensation.

Analysing the behaviour of $E^*$ throughout the trajectories represented in Figure \ref{fig:ene-2step}~A reveals the presence of one additional step change in $E^*$, marking the emergence of a crystal phase within the DLD.
Crystallisation occurred in approximately 1\% of simulations, with long-range order consistently emerging within the DLDs, indicative of a two-step crystallisation pathway. 
This behaviour is not unexpected. Indeed, two-step crystallisation was identified both in simulations and experiments in a range of systems, demonstrating that this pathway to crystals is more prolific than once assumed.\cite{Sosso2016CrystalSimulations,DeYoreo2020ANucleation}

In the seminal work of ten Wolde and Frenkel,\cite{Wolde1997EnhancementFluctuations} simulations indicated that colloid crystallisation occurs in dense fluids when the simulation conditions approach those associated with the fluid-fluid critical point.
In their work, the free energy landscape for crystallisation was projected onto a two-dimensional RC characterising the total size of monomer clusters and the size of the crystalline regions in clusters.
Whilst the lowest energy crystallisation pathway evolved with a near-linear correlation in the two RC variables away from the critical point, large amorphous clusters emerge before the onset of crystalline order close to this point.
It is important to note, though, that this \emph{roundabout} pathway to crystals involves a single energy barrier in the 2D RC space and is not necessarily consistent with the observations in this work, where two activated events are involved in the crystallisation of the initial fluid.

\rev{To consider the proximity of the initial system conditions to the fluid-fluid critical point in our model, we performed an additional 15 simulations.
Each of these was prepared using the same random distribution of particles on a sparse fcc lattice with $\rho^*=0.000485$, but where the reduced temperature was $T^*=1.8-2.2$.
From the densities of the emerging DLDs and nanocrystals in (pseudo-)equilibrium with vapour phases \SI{(see SM Section S2 for details)}, we constructed a $T^*-\rho^*$ phase diagram, shown in  \SI{SM Figure S1}.
This indicates that the simulations in this study at $T^*=2$ are initiated in the immiscible region of the phase diagram and close to the (upper) vapour-fluid critical temperature, $T_c^* \approx 2.05$; hence, $T^*/T_c^* \approx 0.976$.}

As for the nucleation of the DLD, the times associated with the nucleation of a crystalline domain within the DLD are stochastically distributed and are marked by a significant variation in $E^*$ (see Figure \ref{fig:ene-2step}~A). While an escape probability could be built based on the time taken to observe such a sudden change in $E^*$, given the limited statistics, alternative methods to evaluate crystal nucleation times are necessary. 
They will be discussed in the sections below.

Another noteworthy observation from the crystallizing trajectories is that despite crystallization conditions being consistent throughout the entire set of simulations, the structures spontaneously emerging from crystal nucleation appear to differ.
In Figure \ref{fig:ene-2step}~B, the density of energy states representing the crystal in equilibrium with a low-density vapour phase are misaligned in different simulations.
Some systems have a much lower average potential energy than others, despite the crystal phase emerging relatively early on in the trajectories.
These different crystals, characterised by different potential energy levels, result from stacking faults introduced during the rapid propagation of order in the DLD.

By performing polyhedral template matching (PTM),\cite{Larsen2016RobustMatching} we can estimate the numbers of colloids in the single crystals with face-centred cubic (fcc), hexagonal close-packed (hcp), body-centred cubic (bcc) and icosahedral (ico) local symmetries.
Considering the latter stages of trajectories, PTM points to nanocrystals rich in fcc and hcp local environments, while negligible levels of bcc are found.
Figure \ref{fig:ene-2step}~C shows the relative hcp/fcc content as a function of the system potential energy (which at equilibrium is dominated by the potential energy of the crystal).
Crystals with lower hcp/fcc content have the most negative potential energy.
Figure \ref{fig:ene-2step}~D provides example snapshots for three crystals with system potential energies provided inset.
These crystals contain large domains of fcc and hcp particles, though, as is clear in \ref{fig:ene-2step}~D ii and iii, defects are apparent.
Crystal i displays planar fcc and hcp domains with a large hcp core, while crystal ii displays a five-fold hcp symmetric axis with hcp protrusions encompassing fcc domains.

The lowest energy crystals form a cluster in the data in Figure \ref{fig:ene-2step}~C at the more negative end of $E^*$.
None of these crystals contain icosahedral particles.
A higher energy cluster of points centred around $E^* \approx -13.5 \epsilon$, however, include crystals, all of which contain two to three icosahedral particles.
Figure \ref{fig:ene-2step}~D iii provides an example structure where three icosahedral particles introduce a trifold symmetry in the crystal structure.
This motif was a common feature of the crystals in this cluster and seemed to minimise the extent to which fcc and hcp domains grow.
For example, 31\% of particles were identified as fcc or hcp in the lowest energy crystals, while this was 18\% in the higher energy clusters, on average.
A more poorly crystalline structure was found for the system where $E^* = -11 \epsilon$ and four icosahedral particles emerge in the solid, associated with a very limited propagation of the crystal lattice: 10\% of particles in this system can be recognised as matching a crystal structure at the end of the trajectory, and these were nearly all hcp-like.

Different types of defects emerging consistently possibly suggest similar growth patterns for crystals in the DLDs.
The force field used in this work was chosen for efficiency purposes: in order to sample multiple crystallising trajectories, crystallisation must occur over a reasonable simulation timescale.
The result, however, is that even in very small crystals, relaxation of the crystal structure does not readily occur; hence, the defects are locked into the final crystal in the steady state that was sampled.

\subsection{\label{sec:vamp}Crystallisation CVs and VAMP}

In the previous section, $E^*$ and the number of crystal-like particles evaluated using PTM were used to describe phase separation and different crystal structures.
These, however, are only some of the possible CVs that can be used to monitor and describe the crystallisation process.
As described in Section \ref{sec:methods}, we computed a total of 19 CVs, listed in Table \ref{tab:table1}, that may provide good metrics to monitor the evolution of a crystallising system, such as the one adopted in this work.

The concatenated CV time series obtained from 11 crystallising trajectories are shown in \SI{SM Figure S2}. In addition, the CV histograms for one of the trajectories are provided in \SI{SM Figure S3}.
The two-step nucleation process is clearly identifiable in cn.mean, ncl, Q4.mean, laQ4.mean ent, nlcust1 and ene CVs, where step changes in these variables separate time windows where the data are approximately constant within noise.
Some CVs are better suited to identify crystal phases from amorphous ones, and these include ncs, ncnq4, ncnq6, q6.mean, non, fcc and hcp CVs.
Other CVs are best suited to identify condensed phases from vapour phases, such as q4.mean, Q6.mean and laQ6.mean.
Finally, CVs which don't clearly differentiate the probability distributions of states between at least two phases observed in the trajectories are bcc and ico.

\begin{figure*}
\includegraphics[width=0.75\linewidth]{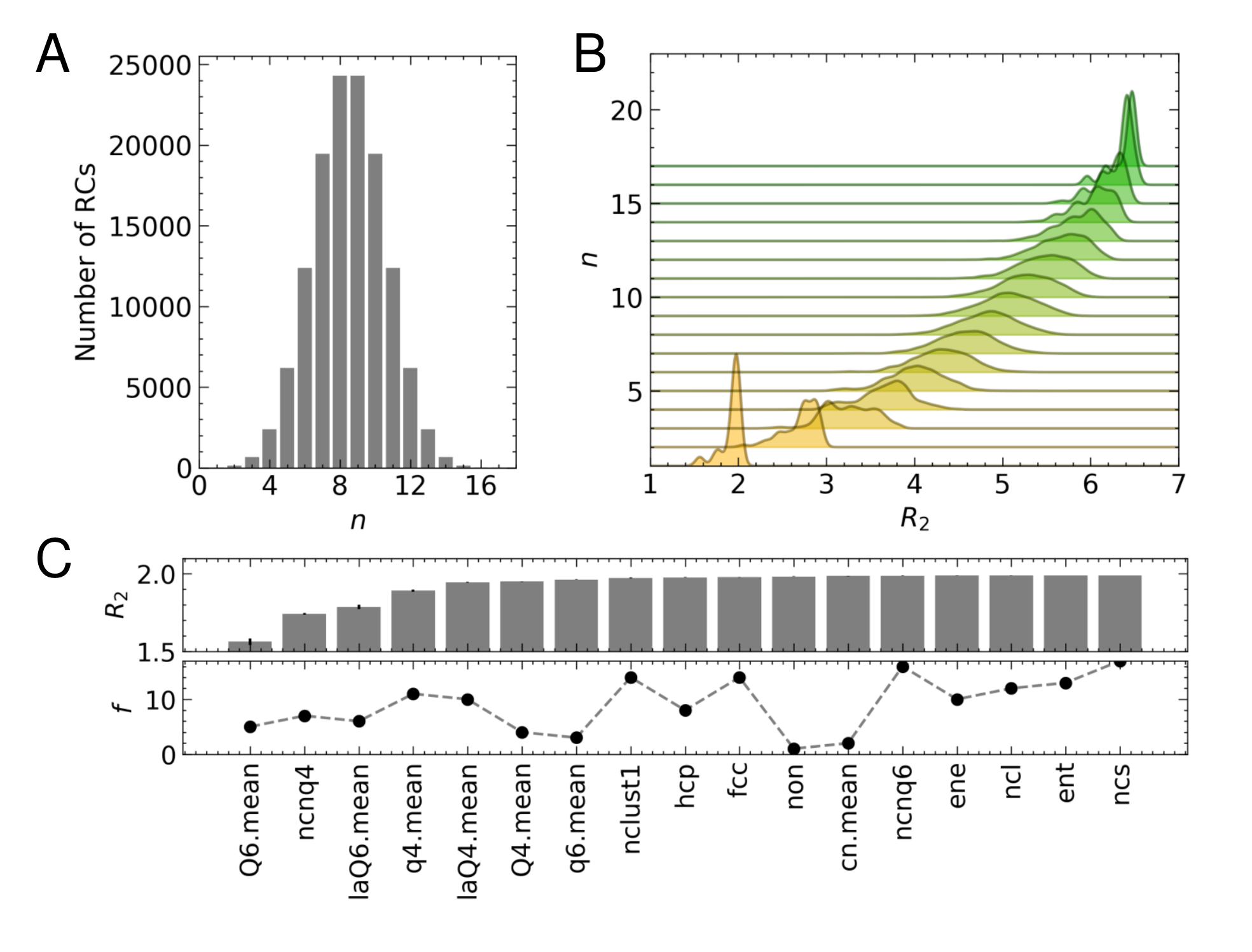}
\caption{\label{fig:vamp} A) Total number of RCs, i.e., combinations of the CVs for every RC dimension, $n$, ranging from 1 to 17. B) Probability densities for the $R_2$ scores for all RCs according to their dimensionality on the $y$-axis. The distributions were evaluated using Gaussian kernel density estimation with a bandwidth of 0.05. C) Top: $R_2$ scores for RCs constructed from one CV. Error bars highlight uncertainties in the scores. Bottom: The number of CV occurrences, $f$, in the highest ranking RCs when $n=1-17$.}
\end{figure*}

In order to determine which CVs best describe the crystallisation dynamics, we constructed RCs containing all possible combinations of CVs as well as those containing a single CV.
The number of possible RCs is given by $2^{\mathrm{max}(n)}-1$, where $n$ is the number of CVs and, therefore, the maximum number of dimensions in any RC.
For this analysis, we did not include the ico and bcc CVs; hence, the maximum $n$ was 17, providing a total of 131,071 RCs.
Figure \ref{fig:vamp}~A provides a histogram for the number of RCs according to the dimensionality of the order parameter space.
Using VAMP, $R_2$ scores were evaluated for each of these RCs with a lag-time, $\tau=20,000$ simulation steps (this equates to a simulation time, $t^*=100$).
Analysis of a range of $\tau$ values indicated that the $R_2$ scores were relatively insensitive to the choice of $\tau$ up to around $\tau=50,000$.
We also chose not to limit the total number of dynamic processes for the given $\tau$; hence, all eigenvalues are used to compute the VAMP-2 scores.
In this analysis, the trajectory which led to a poorly crystalline solid was neglected, and the $\mathbf{f}(\mathbf{r},t)$ arrays were constructed using absolute, normalised CV values, such that CVs range from zero to one.
The distributions for these rescaled CVs from the combined trajectories are provided in \SI{SM Figure S4}.

Figure \ref{fig:vamp}~B provides the distributions of $R_2$ scores for all of the RCs.
For monodimensional RCs ($n=1$), the $R_2$ scores are provided for each CV in Figure \ref{fig:vamp}~C (top panel).
All $R_2$ scores are greater than the minimum of one, which would indicate invariant sampling of the RC. The best scoring CV is ncs with $R_2=1.991 \pm 0.001$; within statistical uncertainties, however, ncs, cn.mean, ncl, ncnq6, ent and ene are equal.
Not all of these CVs were identified as best suited to follow the two-step mechanism, but they all identify the emergence of a crystalline particle.
Apart from cn.mean, ene and ent, the ten highest-scoring CVs are determined by counting the number of particles according to the density or symmetry of their local coordination environment.
When a monodimensional RC is used to study crystal nucleation, such as in CNT-based seeding methods,\cite{Knott2012HomogeneousConditions,Sanz2013HomogeneousSimulation,Zimmermann2015NucleationRates} CVs quantifying the size of the emerging phase based on local structure are adopted.
Our analysis here validates that these features (i.e., total numbers of particles with solid-like first-sphere coordination numbers or particles with high local coordination symmetries reminiscent of the crystal) are good indicators for the slow dynamics of the system.
This is consistent with the results from likelihood maximisation (and validated using committor analysis), which identified that the best 1D reaction coordinate to study crystallisation in Lennard-Jonesium liquid was a product of the nucleus size and the local $Q_6$ CVs. \cite{Beckham2011OptimizingDuration}
Monodimensional RCs based on fourth-order Steinhardt parameters, as well as Q6.mean, in our work, are low-ranking indicators for crystallisation; this is perhaps unsurprising in the case of $Q_4$-based CVs, given that crystals display fcc and hcp particle packing.

As discussed in Section \ref{sec:introduction}, some simulation studies of crystallisation adopt RCs constructed from two CVs to investigate pathways in systems where crystalline order emerges from amorphous clusters. \cite{Wolde1997EnhancementFluctuations,Jiang2019NucleationSpinodal,Bulutoglu2022AnClusters,Salvalaglio2015UreaDependent,Salvalaglio2015Molecular-dynamicsSolution,Finney2022MultipleNucleation}
In such cases, the RCs characterise cluster size/density and relative cluster crystalline order in orthogonal degrees of freedom to evaluate pathways from supersaturated solutions to crystals.
Provided this context, we consider 2D combinations of CVs that rank highly in the VAMP analysis.
Given the 136 possible combinations of CVs used to propose a two-dimensional RC candidate here, three scored equally highly; these were, \{ncl, ncnq6\}, \{ncs, ncnq6\} and \{ncnq6, ene\} where $\overline{R_c} = 2.939 \pm 0.006$. 
These were followed by a second tier set with $\overline{R_c} = 2.929 \pm 0.011$: \{ncnq6, nclust1\}, \{ncl, fcc\}, \{cn.mean, ncnq6\}, \{ene, fcc\}, \{cn.mean, fcc\}, \{ncnq6, ent\} and \{ent, non\}. 

Generally, the highest ranking 2D RCs combine CVs, one of which distinguishes well the two-step pathway and another which clearly identifies the emergence of crystalline order.
It is notable that the ncnq6 variable appears in 6 of the 10 highest-scoring RCs.
This is the only CV that is zero in the absence of a crystalline phase and perfectly resolves any degeneracy between disordered and ordered clusters. 
As in the case of the monodimensional RCs, many of the CVs listed above scale with the size of emerging phases.
In our previous work on NaCl crystallisation, we adopted an RC using two CVs to characterise the size of dense ion clusters and the level of crystalline order in these regions to follow crystallisation where multiple pathways to crystals are evident, including those where order emerges in liquid-like intermediates. \cite{Finney2022MultipleNucleation}
The closest RC analogue in the current work to the one adopted previously is \{ncl, ncnq6\}, which is among the highest-scoring set of 2D CVs and indicates that, for the specific problem at hand, a choice driven by observation and intuition was able to identify a good set of candidate CVs.

Across the entire range of $n$, adding more descriptors for collective particle features leads to shifting of the $R_2$ distributions to higher values (see Figure \ref{fig:vamp}~B): $\log{(\tilde{R_2})} = -0.03 (\log{n})^2+0.46\log{n} + 0.3$, where $\tilde{R_2}$ indicates the median, and the coefficient determining the fit is $0.99$.
In the case of the highest ranking RCs, $R_2$ converges to a maximum around 6.5 when $n=15-17$; here, $\log{(R_2)} = -0.3 (\log{n})^2+0.79\log{n} + 0.3$.
Thus, adding more descriptors for crystallisation increases the VAMP-2 score and provides RCs that more accurately capture the slow modes.
Given the small increases to $\max{(R_2)}$, however, for large values of $n$, it is possible to trade off computational efficiency with the accuracy to determine the kinetics for these transitions. 

The best performing RCs when $n=1-17$ tend to comprise CVs such as ncs (i.e. CVs which identify the size of crystalline regions) as well as ene and ent (see \SI{SM Table S2}).
Indeed, the highest-scoring CVs in monodimensional RCs feature in the highest-ranking multidimensional RCs, as shown in Figure \ref{fig:vamp}~C (bottom panel).
While this observation is general, there are notable exceptions in the case of cn.mean and non.
This is perhaps not surprising, given that the time-dependent ncs and ncl values are highly correlated with cn.mean. Similarly, fcc and hcp time series are highly correlated with non.
Generally, though, analysis of the full spectrum of possible CV combinations identified some CVs as better than others at monitoring a two-step crystallising system.
Our analysis supports the conclusions from previous simulation studies demonstrating that CVs which better characterise the \emph{local} symmetry in the first-coordination sphere and those that quantify the size of emerging phases are the best candidates to describe and follow the crystallisation process. \cite{Zimmermann2015NucleationRates,Lechner2008AccurateParameters,Pretti2019Size-dependentCrystallization,Savage2009ExperimentalCrystallization,Fang2020Two-stepMixtures}

In the following sections, we further assess the performance of RCs obtained by combining different CVs by constructing Markov State Models and using them to compute nucleation rates, mechanisms, and associated free energies of the relevant (meta)stable states.

\subsection{Markov State Models}

With knowledge of the VAMP-2 scores, it is interesting to see how the rates for crystal nucleation compare when evaluated using RCs constructed from the highest-scoring CV combinations. 
In this section, we build MSMs for all of the highest-scoring RCs for $n=1$ to $n=17$ to identify (meta)stable states and transitions between them.
Furthermore, in order to compare the system representation across RCs with different dimensionality, we use TICA to project the high-dimension RCs onto just two coordinates that best separate the states of interest.
As TICA quantifies the variance in the crystallisation kinetics, the dimensionality reduction produces a reaction coordinate where the distance between (meta)stable states is a function of the system's time evolution.\cite{Perez-Hernandez2013IdentificationConstruction}

\begin{figure*}
\includegraphics[width=0.75\linewidth]{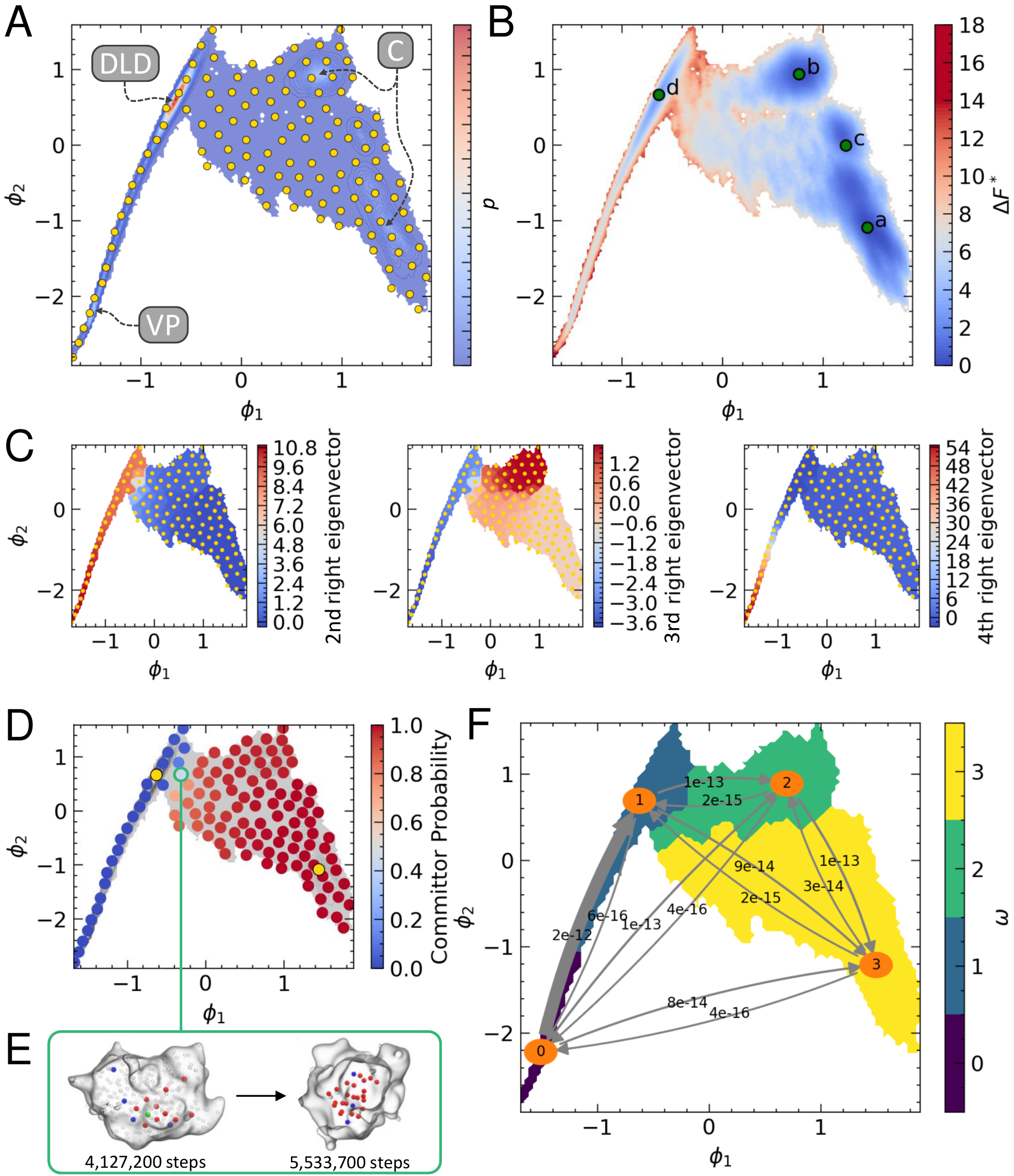}
\caption{\label{fig:17CVMSM} Results from a Bayesian MSM constructed from 17 CVs evaluated for 15 independent simulation trajectories projected onto two TICA coordinates: $\phi_1$ and $\phi_2$. A) The relative probability density (highlighted by the colour scale) of the sampled states in the 2D TICA RC, with (meta)stable vapour phase (VP), dense liquid droplet (DLD) and crystal peaks (C) indicated by the arrows. The positions for 119 partition centres used in the construction of the MSMs are overlaid and shown as yellow circles. B) Relative free energies ($\Delta F^*$) in units of $k_\mathrm{B}T^*$ computed from the Bayesian weighted stationary distribution projected onto the sampled states; labels a, b, c and d identify the four lowest energy minima in the landscape when determined using the approach discussed in the text. C) Projections of right eigenvectors $2-4$ onto the sampled states and with partitions also highlighted. D) Committor probabilities, highlighting the probabilities for partitions to commit to either the DLD or C$_3$ basins indicated by the yellow circles. E) Snapshots of microstates from a single trajectory which are associated with the partition in D where the committor probability is $\approx 0.5$ (see Figure \ref{fig:ene-2step} caption for a description of the representation).
F) Transitions between the four PCCA+ (meta)stable states: VP (0), DLD (1), C$_2$ (3) and C$_4$ (4). The labels indicate the rates, also provided in Table \ref{tab:table2}, and the width of the arrows indicates the fastest transitions.}
\end{figure*}

\subsubsection{MSM from $n=17$ RC data} 
We begin by discussing the general features arising from a Bayesian MSM constructed using the highest-ranking RCs following the protocol below.
In this case, the RC comprised 17 CVs; the best performing RCs when $n=14-17$ had the same $R_2$ score within statistical uncertainties.
As for the calculation of $R_2$ values, we used the absolute, scaled CV coordinates to build the MSM.
To reduce the uncertainty in the estimate of the slowest timescales, following an initial analysis, we complemented the set of 10 reactive trajectories discussed in the previous sections with an additional five simulations, four of which produced a DLD and another which led to a crystal with mean $E^* = -17.282 \pm 0.157 \epsilon$.
The addition of these data resulted in no qualitative differences in the MSMs but did facilitate a more accurate determination of crystallisation kinetics---the slowest implied timescales were $1.078 - 2.198 \times 10^7$ steps and $1.164 - 1.789 \times 10^7$ steps within a 95\% confidence interval (CI) before and after including the additional simulation trajectories. Note that in what follows, we report the timescales and rates in terms of numbers of trajectory steps, with each step corresponding to a reduced time of 0.005.

First, we applied TICA to project the time-dependent 17 CV values from 15 simulation trajectories onto the \{$\phi_1$,$\phi_2$\} TICA RC where $\tau=10,000$ steps in the evaluation of the time-lagged components. Though the calculation of VAMP-2 scores was rather insensitive to the choice of $\tau$ when $\tau<50,000$ steps, identifying the fastest processes in the system dynamics requires a smaller value of the lagtime (as discussed below, the implied timescales from the model shown in \SI{SM Figure S6} indicate that this was a reasonable choice to distinguish all of the relevant transitions).
The concatenated $\phi_1$ and $\phi_2$ trajectories resulting from TICA are provided in \SI{SM Figure S5}. These indicate that $\phi_2$ clearly separates different crystal states resulting from crystallisation. On the other hand, $\phi_1$ shows distinct time windows where dense amorphous phases i.e., DLDs, are present.

Figure \ref{fig:17CVMSM}~A shows the cumulative sampled probability density of states as a function of \{$\phi_1$,$\phi_2$\}. 
A small peak for states in the VP is observed at $\phi_1 \approx -1.4$, $\phi_2 \approx -1.8$, while a much more pronounced peak at $\phi_1 \approx -0.6$, $\phi_2 \approx 0.6$ accounts for microstates in the DLD.
Two to three broad peaks highlighted on the plot are observed for states where crystals are present.
The wide distribution of the states highlights the slow time evolution of the crystals during the simulations.

To construct the MSM, the sampled configurations were mapped onto a discrete set of partitions in the \{$\phi_1$,$\phi_2$\} space using regular space clustering with a minimum distance of $\phi =0.2$. 
The attribution of microstates to a partition was carried out by Voronoi tessellation of the sampled data\cite{Prinz2011MarkovValidation}. Partition centres are shown in Figure \ref{fig:17CVMSM}~A. 
This procedure generates trajectories describing transitions between discrete partitions, which can be used to construct a transition matrix. We confirmed that the resulting 119 partitions were fully connected in the MSM, and all transitions between partitions were used to determine kinetic information from the fully connected network of partitions.

The probability density weights associated with partitions determine the stationary distribution of states, which can be Boltzmann-inverted to generate the free energy landscape in \{$\phi_1$,$\phi_2$\}, provided in Figure \ref{fig:17CVMSM}~B.
The landscape indicates a narrow reactive pathway associated with the VP to DLD transition, corresponding to the condensation process. Instead, the path from the DLD to different crystal states is less constrained in the RC space.
The model accurately determines the relative stability of the different crystalline nuclei observed in simulations
. Crystals with the lowest potential energy, in fact, correspond to the global minimum in the free energy landscape (see point a in Figure \ref{fig:17CVMSM}~B).
The two metastable states corresponding to local minima of the free energy (determined using moving $20 \times 20$ windows in a $150 \times 150$ grid of the RC space) and identified by labels b and c, also represent crystal nuclei associated with a $\Delta F^* = 0.26$ and $1.17 \; k_\mathrm{B}T^*$, respectively.
The difference in the free energies between a DLD and the most stable crystal,  $\Delta F^* \approx 3 \; k_\mathrm{B}T^*$ and, though not shown in the Figure, $\Delta F^*$ for the basin representing VP microstates is $6  \; k_\mathrm{B}T^*$ (the energies are shifted so that at the global minimum, $F^*=0$).
The ranking of relative stabilities of the VP, DLD, and crystals, which can be qualitatively inferred by observing the dynamic trajectories, is therefore captured well by the MSM.

The eigenfunctions of the MSM approximate the transitions between (meta)stable states in the system.
The eigenvalues associated with these functions determine their importance when predicting the time evolution of the system.
\SI{SM Figure 6} provides the implied timescales for the twelve slowest transitions as a function of different $\tau$ values, which are computed as $t_i = -\tau / \ln{|\lambda_i(\tau)|}$ and where $\lambda_i$ is the eigenvalue for process $i$.
These implied timescales all increase as a function of $\tau$ but plateau when $\tau \approx 100$ steps; indeed, this analysis was used to identify the appropriate value of $\tau$ in a series of trial and improvement cycles.
Several orders of magnitude separate the implied timescales for the slowest and fastest transitions, which is not surprising given the reaction under consideration.

By projecting the right eigenfunctions with the largest eigenvalues onto the TICA RC, we can visualise the slowest modes in the system. Figure \ref{fig:17CVMSM}~C shows that the slowest transition is from crystalline partitions to the VP; the second slowest process is one which goes from the VP to higher energy crystals via the lowest energy crystal partitions; while the third slowest process is a transition from condensed phases to a VP.
It is important to note that, though we never observe these transitions in the forward reactive trajectories generated in the simulations, the construction of the MSM through the partitioning of states in energy minima and transition state regions means that we can predict these slow modes.
Provided the free energy landscape in Figure \ref{fig:17CVMSM}~B and physical intuition, these slowest transitions are to be expected with the model assumption of ergodicity.
To test the accuracy of the model predictions, we performed a Chapman-Kolmogorov test \cite{Prinz2011MarkovValidation} using four (meta)stable states, the results for which are shown in \SI{SM Figure S7}. 
This test evaluates the left- and right-hand sides of the equation $\mathbf{T}(k \tau) = \mathbf{T}^k (\tau)$, where $\mathbf{T}$ is the transition matrix, and $k$ is the number of trajectory steps we adopt in the calculation.
The results in \SI{SM Figure S7} indicate that the model predictions and estimates from the data are consistent.

The 2$^\mathrm{nd}$ and 3$^\mathrm{rd}$ eigenvectors in Figure \ref{fig:17CVMSM}~B highlight the approximate transition between amorphous and crystalline states.
To explore this more accurately for the forward transition associated with the onset of crystalline order within the DLD, we performed a committor probability analysis considering the DLD and the most stable crystal minimum as end states, as shown by the yellow circles in Figure \ref{fig:17CVMSM}~D.
The partitions in the Figure are coloured blue to red according to their probability of committing to the crystal basin. 
The transition state ensemble projection onto \{$\phi_1$,$\phi_2$\} corresponds to the region of CV space approximately identifying the isocommittor. 
In particular, the partition highlighted by a green circle in Figure \ref{fig:17CVMSM}~D, has a committor probability of 0.5, providing the closest approximation of the transition state (TS) associated with the crystal nucleation transition. Figure \ref{fig:17CVMSM}~E provides snapshots of the dense phase at the beginning and end of a portion of a single trajectory crossing the TS partition, where the crystal-like particles are identified using PTM.
It is clear that the number and local density of the particles with crystal-like local environments increases (highlighted by the colours in Figure \ref{fig:17CVMSM}~E). Moreover, their arrangement appears to become more ordered, in line with what is expected for the second step in a two-step crystallisation mechanism.

\begin{table}
\caption{\label{tab:table2}Transitions between the (meta)stable states in a 2D TICA RC constructed using 17 CVs, ranked according to their rates calculated from MFPTs. 95\% confidence intervals (CI) in the rates are also provided and the units are ($\sigma^3$ steps)$^{-1}$. }
\begin{ruledtabular}
\begin{tabular}{llcr}
 & Transition & Rate & 95\% CI\\
\hline
1 & VP $\rightarrow$ DLD & $1.85 \times 10^{-12}$ & $1.47 - 2.36 \times 10^{-12}$\\
2 & C$_2$ $\rightarrow$ C$_3$ & $1.18 \times 10^{-13}$ & $0.87 - 1.58 \times 10^{-13}$\\
3 & DLD $\rightarrow$ C$_2$ & $1.13 \times 10^{-13}$ & $0.85 - 1.43 \times 10^{-13}$\\
4 & VP $\rightarrow$ C$_2$ & $9.59 \times 10^{-14}$ & $0.77 - 1.16 \times 10^{-13}$\\
5 & DLD $\rightarrow$ C$_3$ & $8.58 \times 10^{-14}$ & $0.68 - 1.05 \times 10^{-13}$ \\
6 & VP $\rightarrow$ C$_3$ & $7.55 \times 10^{-14}$ & $6.12 - 9.07 \times 10^{-14}$\\
7 & C$_3$ $\rightarrow$ C$_2$ & $3.41 \times 10^{-14}$ & $2.47 - 5.26 \times 10^{-14}$\\
8 & C$_2$ $\rightarrow$ DLD & $1.77 \times 10^{-15}$  & $1.08 - 2.67 \times 10^{-15}$ \\
9 & C$_3$ $\rightarrow$ DLD & $1.72 \times 10^{-15}$ & $1.06 - 2.57 \times 10^{-15}$\\
10 & DLD $\rightarrow$ VP & $5.92 \times 10^{-16}$ & $3.76 - 9.22 \times 10^{-16}$\\
11 & C$_2$ $\rightarrow$ VP & $3.98 \times 10^{-16}$ & $2.64 - 5.75 \times 10^{-16}$\\
12 & C$_3$ $\rightarrow$ VP & $3.96 \times 10^{-16}$ & $2.62 - 5.71 \times 10^{-16}$\\
\end{tabular}
\end{ruledtabular}
\end{table}

To determine the rates in two-step crystallisation, we performed a spectral clustering of the partitions using Robust Perron Cluster Cluster Analysis (PCCA+)\cite{Roblitz2013FuzzyClassification} to cluster partitions according to the eigenvectors of the transition matrix associated with the MSM; \SI{SM Figure S8} highlights the weights for each partition assignment to states.
The assignment of partitions to four (meta)stable states, $\omega$, using this approach is shown in Figure \ref{fig:17CVMSM}~F, and the fraction of microstates associated with $\omega=0$, 1, 2 and 3 was 0.0004, 0.0104, 0.2487 and 0.7405, respectively.
These states represent the VP, DLD, C$_2$ and C$_3$ in order of increasing $\omega$, where C$_2$ are crystals with higher potential energy and C$_3$ are the more stable crystals.

The mean first passage times (MFPTs) between (meta)stable states can be determined from the transition matrix of the MSM.
Rates computed from these MFPTs and their uncertainties, determined within a 95\% CI are provided in Table \ref{tab:table2}, which indicate that the fastest transition is the condensation of the VP to form a DLD.
The fastest transitions following this are the emergence of order in the DLD to form higher energy crystals (C$_2$) and the transformation of C$_2$ to C$_3$ crystals.
We did not observe this latter transition during the simulations, as already discussed, and so the quantitative predictions of the model here should be further tested.
The slowest transitions are those already identified as transformations of crystals to the VP and DLD phases.
Faster transitions occur from the VP to crystals; however, the distribution of states indicates that the system must first go via the DLD.
Indeed, following the forward reaction, the MSM indicates that the crystallisation pathway proceeds according to VP $\rightarrow$ DLD $\rightarrow$ C$_2$ $\rightarrow$ C$_3$, and, in general, the predictions of the MSM are consistent with the pathways and relative kinetics of transitions that are to be expected for a two-step crystallising system.

\subsubsection{MSMs for $n=1-17$ RCs} 
The approach laid out above for $n=17$ can be applied to describe the mechanisms and compute the crystallisation rates with other combinations of CVs.
We, therefore, constructed MSMs for all of the $n=2- 16$ highest $R_2$ scoring CV combinations.
To ensure a fair comparison of model results, minimal changes were made during the construction of MSMs; hence, we first projected the CVs onto two TICA coordinates using $\tau=10,000$ steps and computed the stationary distributions and transition matrices by sampling discrete partitions in a TICA 2D RC space.
As before, we ensured that the value of $\tau$ and the partitioning of states led to converged implied timescales for the slowest modes, along with fully connected partitions, as well as model predictions for transitions which were consistent with sampled data in Chapman-Kolmogorov tests of the constructed MSMs.

In general, the MSMs when $n=2-16$ identified the same qualitative and quantitative features identified and discussed for $n=17$.
Some notable differences were that the relative free energy differences between the (meta)stable states fluctuate within $\sim 2 k_\mathrm{B}T^*$, particularly for smaller values of $n$. Nevertheless, when $n=6-17$, $\Delta F^*$ for the DLD to crystal transition converges to $-3.1 \pm 0.3 \; k_\mathrm{B}T^*$.
For our purposes, however, amorphous phases were always higher in energy than crystalline states, and the forward reaction i.e., VP$\rightarrow$crystal, was always predicted to be significantly faster than the reverse reaction in all of the MSMs. 
In addition, while there was some reordering of the implied timescales for state-to-state transitions, the relative ranking of the transitions involved in the formation of crystals from the VP was consistent throughout, regardless of the choice of CVs used to construct the RC.
This is a good sign of the robustness of the kinetic models to capture the slow transitions with reasonable choices for the CVs that can describe the time-dependent structural evolution of the system.

\SI{SM Figure S9} provides the state maps evaluated when $n=2-17$ from each Bayesian MSM and PCCA+ to identify the (meta)stable states.
These reflect the small change in the assignment of partitions to (meta)stable states and the order of state-to-state transitions when $n>5$.
As $n$ decreases, the extent of$\phi_1$ and $\phi_2$ in the 2D TICA RC tend to increase, resulting in more partitions of the sampled data.
Despite this, the fraction of partitions assigned to the VP state decreased. It was necessary  to increase the minimum distance between partitions from $\phi = 0.2$ to $\phi = 0.22$ in the regularly spaced clustering algorithm when $n=2-5$.
As well as a reduction in the number of partitions for the VP phase, expansion of the TICA coordinate range is concomitant with a broadening of the crystal state regions (see the relative areas for $\omega_0$/$\omega_1$ and $\omega_2$/$\omega_3$ in \SI{SM Figure S9}).
Another interesting feature is that when $n \leq 4$, \SI{SM Figure S9} shows that the TICA projection of states is reflected in $\phi_2$, such that VP microstates are found when $\phi_2$ is at positive values, unlike in Figure \ref{fig:17CVMSM}.

In the case of $n=2$, the area in the state map for the VP and DLD is very small, and the number of partitions representing these states is substantially decreased \textit{cf.} $n=17$ (though the fraction of states in amorphous phases remains constant at around 0.011).
Due to the small number of partitions, particularly in the VP region, spectral clustering can capture only three (meta)stable states, where $\omega_0$, in this case, includes all of the non-crystalline microstates.
This model was constructed using a 2D TICA projection of \{ncs,ncnq6\} where both CVs are designed to identify the emergence of crystals.

Equally highly scoring $R_2$ 2D RCs were \{ncl, ncnq6\} and \{ene, ncnq6\}; hence, we constructed Bayesian MSMs using 2D TICA projections of these sampled CVs.
As shown in \SI{SM Figure S10}, the RC constructed from ncl and ene (along with ncnq6) are very similar to those constructed from the highest-scoring CV pair.
Applying PCCA+ to the partitions, however, does result in four (meta)stable states with a separation of the VP and DLD, and with a forward transition between amorphous states on the order of $1 \times 10^{12}$ $(\sigma^3 \; \mathrm{steps})^{-1}$, consistent with MSMs built from RCs capturing greater numbers of degrees of freedom.
Here, the rates for crystal nucleation in the DLD were $0.98 - 1.87 \times 10^{13}$ and $1.13 - 1.89 \times 10^{13}$ $(\sigma^3 \; \mathrm{steps})^{-1}$, respectively, with 95\% statistical confidence in the values, compared to the rate of $0.78 - 1.39 \times 10^{13}$ $(\sigma^3 \; \mathrm{steps})^{-1}$ predicted in the case of \{ncs, ncnq6\}.
Though there is overlap in the rate predictions, \{ncs, ncnq6\} results in a slower nucleation rate.

It is possible to compute a VAMP-2 score for the time series of TICA coordinates which we label $R_2^{TICA}$.
The $R_2^{TICA}$ for the three reaction coordinates \{ncl, ncnq6\}, \{ncs, ncnq6\} and \{ene, ncnq6\}, were $1.964 \pm 0.002$, $1.961 \pm 0.005$ and $1.96 \pm 0.002$, respectively, reflecting the earlier observation that all CV combinations are able to capture the slow variations in the underlying system dynamics.
Despite this, the mechanistic insight provided by the three models differs, and this is somewhat sensitive to the method used to discretise trajectories and identify (meta)stable states.
Care should be taken, therefore, when assessing model outcomes.

\begin{figure*}
\includegraphics[width=0.75\linewidth]{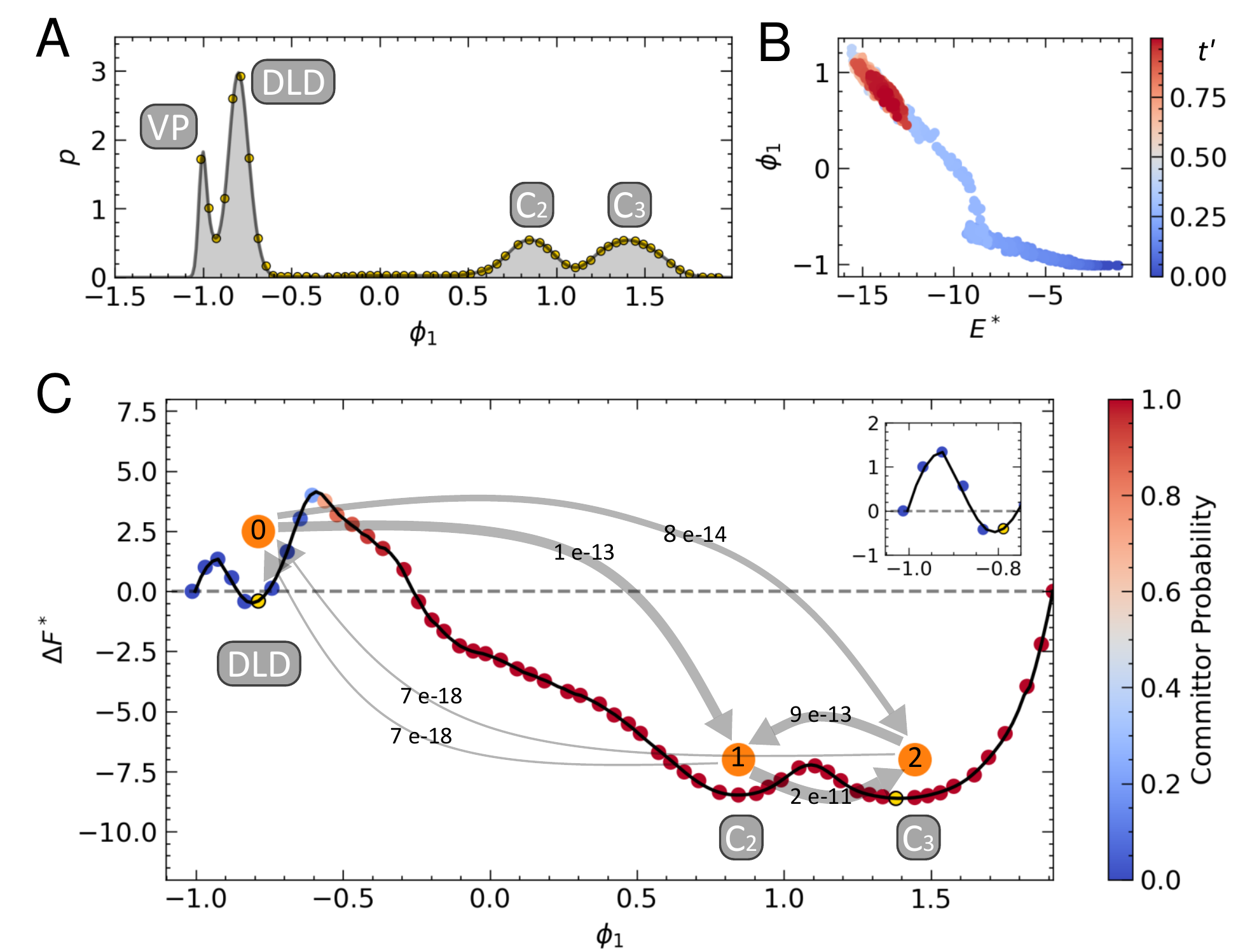}
\caption{\label{fig:1DMSM} Results from a Bayesian MSM constructed from the ncs CV evaluated for 15 independent simulation trajectories and projected onto a TICA coordinate, $\phi_1$. A) Probability densities for $\phi_1$ states with partition centres used in the MSM indicated by the yellow circles. Peaks indicate the (meta)stable states which are labelled. B) A single trajectory plotted as a function of ene and $\phi_1$; the colour scale indicates the scaled simulation time. C) Relative free energy (in units of $k_\mathrm{B}T^*$) as a function of $\phi_1$ with the inset frame showing $\Delta F^*$ at small values of $\phi_1$. The circles indicate the partition centres, coloured according to their probability to commit to the yellow partition centre in the C$_3$ minimum from the DLD minimum. States identified using PCCA+ are labelled 0, 1 and 2, with arrows between the states indicating the rate for transitions computed using MFPTs.}
\end{figure*}

In the case of $n=1$---ncs provided the highest $R_2$ scoring CV---only one TICA coordinate ($\phi_1$) was used to construct a Bayesian MSM.
For this reason, we used a minimum distance $\phi_1=0.04$ to generate partitions during regular space clustering, resulting in 59 partitions.
The probability density of states in $\phi_1$ is provided in Figure \ref{fig:1DMSM}~A, with the VP, DLD, C$_2$ and C$_3$ crystal states clearly apparent in the TICA projection of ncs.
Figure \ref{fig:1DMSM}~B provides the time-dependent trace in \{ene, $\phi_1$\} for a single crystallising trajectory, highlighting how the TICA coordinate values are correlated with CV values.
In the case of ene, there is a clear non-linearity in the data, as was also observed e.g., for \{ncnq6, $\phi_1$\} and \{ent, $\phi_1$\}, while \{ncs, $\phi_1$\} shows a near-perfect linear correlation in the coordinates.
In all of these high $R_2$ scoring CVs, the distribution of TICA coordinates clearly distinguishes crystal and non-crystalline microstates.

Figure \ref{fig:1DMSM}~C provides the free energy profile, determined from a 1D MSM using ncs data, aligned such that $\Delta F^*=0$ for the vapour.
A small energy barrier separates the VP from the DLD, while a more pronounced energy barrier separates the DLD from the crystalline states.
In the latter, there is good agreement between the position of the maximum in $\Delta F^*$ and the partition committor probabilities to commit to either the DLD or C$_3$.
It is clear that the choice of CV affects the relative weights associated with states and, therefore, their $\Delta F^*$ values, since this is a function of the stationary distribution computed using Bayesian MSM weights.
For example, a 1D MSM constructed using a 1D TICA reduction of the ncl CV data provides a free energy profile shown in \SI{SM Figure S11}; here, $\Delta F^*$ between the minimum representing the VP and DLD is around  $2.5 \; k_\mathrm{B}T^*$, compared with $\sim 0.5 \; k_\mathrm{B}T^*$ from Figure \ref{fig:1DMSM}~C.
Qualitatively, the order in the stability of the VP, DLD, C$_2$ and C$_3$ is consistent across the entire $n=1-17$ range, but the $\Delta F^*$ values for DLD$\rightarrow$crystal change from $\sim 1$ to $\sim 8 \; k_\mathrm{B}T^*$ depending on the choice of CVs and the level of reduction in the dimensionality.
Despite this, the free energy difference between the VP and crystals is approximately consistent with the $\Delta F^*$ when additional CVs are included in the constriction of MSMs ($\Delta F^* \approx -8 \; k_\mathrm{B}T^*$ in the monodimensional RC for the forward reaction, compared to $\Delta F^* =-5.8 \pm 0.5 \; k_\mathrm{B}T^*$ when $n=6-17$).

Where the 1D MSM does provide consistent quantitative information with MSMs constructed using additional CV dimensions, is in the overall crystallisation rate. When the highest scoring CVs were used to construct the MSM, we found that the transition rate from the DLD to crystals was around $1 \times 10^{-13} \; (\sigma^3 \; \mathrm{steps})^{-1}$.
As for the $n=2$ case, spectral clustering provided only one amorphous state, centred in the DLD and marked by 0 in Figure \ref{fig:1DMSM}~C, to determine state-to-state transitions---this was a general observation for MSMs constructed for all of the high scoring 1D RCs.
The rates indicate that the C$_2 \rightarrow$C$_3$ transition is the fastest between the three identified (meta)stable states.

\begin{figure*}[ht]
\includegraphics[width=0.75\linewidth]{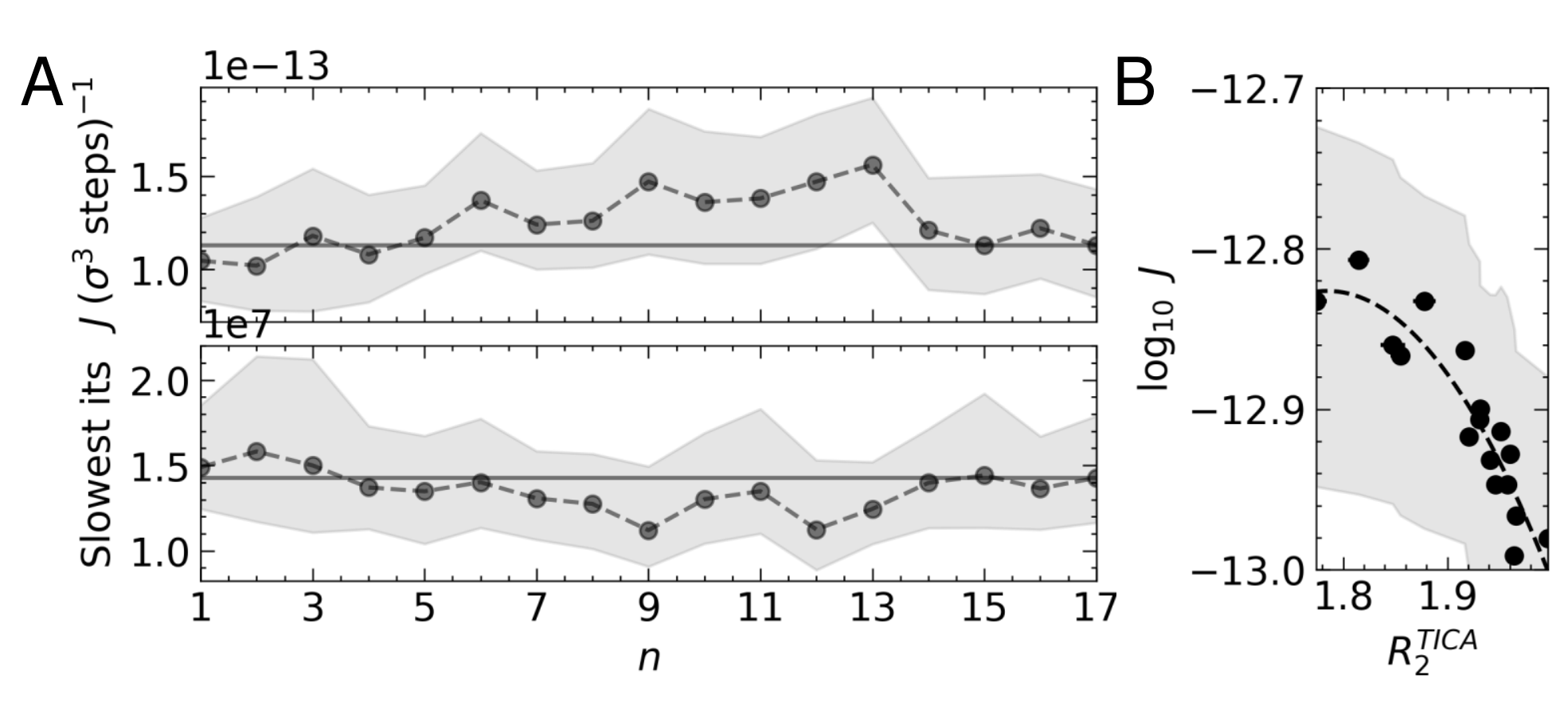}
\caption{\label{fig:n_rates} A) Top: Crystal nucleation rate, $J$, determined from the MFPT for transitions from an amorphous phase to a crystalline one in the Bayesian MSMs as a function of $n$, the number of CVs used to generate the TICA trajectories. Bottom: The slowest implied timescales from the MSM as a function of $n$. Solid lines mark the mean data for the $n=17$ case. B) Logarithm of the rates in A are plotted against the $R_2^{TICA}$ score for the TICA RC. The dashed line is a fit to the data. Shaded regions indicate a 95\% CI in the mean points throughout.}
\end{figure*}

A 1D representation of the free energy pathways to crystals from the supersaturated vapour phase demonstrates how the picture for crystal nucleation differs from one expected for a single-step transformation of the vapour to a crystal following established nucleation theories based on the earliest ideas of Gibbs. \cite{Kashchiev2000Nucleation:Applications}
The two energy barriers may be perceived as a clear departure from a CNT-based model for phase separation; however, one can interpret the two barriers as two distinct steps, each of which can be reasonably well described using CNT-based theories.
The interfacial tension used to predict the crystal nucleation barrier in CNT must account for the crystal lattice's emergence in a DLD.
This phase separation process is consistent with Ostwald's rule of stages,\cite{Ostwald1897StudienKorper} where the first product from nucleation is a thermodynamic phase with chemical potential closest to the parent phase, and which subsequently undergoes further transformations to more stable states.
Furthermore, the pathway is distinct from those where amorphous intermediates do not represent a depression in the free energy landscape. \cite{Kashchiev2020ClassicalCrystals}

Multi-step crystallisation pathways are known experimentally for colloidal systems,\cite{Zhang2007HowCrystallization,Savage2009ExperimentalCrystallization} and pathways to crystals via amorphous intermediates were reported for other crystallising systems. \cite{Vekilov2010TheSolution,DeYoreo2020ANucleation}
These pathways may also include intermediate crystal phases; however, we believe that the different crystal minima in our work are the result of the stacking faults already discussed and not thermodynamically distinct phases at equilibrium which are a feature of, e.g., binary colloid mixtures. \cite{Pretti2019Size-dependentCrystallization,Fang2020Two-stepMixtures}

In all of the MSMs constructed, the slowest process to crystallisation is the second step, i.e., the emergence of order in the liquid.
Figure \ref{fig:n_rates}~A provides the nucleation rates for this step, $J$, which are roughly constant as a function of $n$.
Assuming that the highest dimension CV description of the crystallisation dynamics is the best choice to predict the kinetics as indicated by the higher VAMP-2 score, the most significant departure in the mean rates computed using MFPTs occurs when $n \approx 9-13$.
However, the clear overlap of the 95\% confidence intervals of the rate estimates allows us to confidently determine the rates within the same order of magnitude, indicating that all of the TICA RCs, constructed from a basis of high-scoring CVs of increasing dimensionality, predict consistent nucleation times.

Figure \ref{fig:n_rates}~B shows how DLD$\rightarrow$crystal nucleation rates change as a function of the $R_2^{TICA}$ scores.
The TICA RCs which have a smaller $R_2^{TICA}$ value are in the range $n=9-13$, where $J$ is higher, and the slowest implied timescales in the model (see Figure \ref{fig:n_rates}~A) show a departure from the solid line marking the mean values for $n=17$, while the highest $R_2^{TICA}$ score was for $n=1$.
The dashed line in the Figure is a fit to the data with functional form $\log_{10}(J) = -3.89(R_2^{TICA})^2 + 13.85(R_2^{TICA})+ 25.18$. 
This indicates that the RCs, which best capture the slowest dynamics in the system, also predict slower mean rates for the nucleation of crystals in the DLD. 
It is important to reiterate, however, that the uncertainties mean that the crystallisation rates are predicted consistently in the MSMs, regardless of the CVs chosen to characterise the process and the projection of these onto their time-lagged independent components.

\section{\label{sec:conclusions}Conclusions}

VAMP analysis of the CV time series data from crystallising trajectories indicates that CVs characterising the size of emerging phases in the system often feature in the highest-scoring CV combinations that best describe the slow dynamics for crystallisation.
That these CVs, along with, e.g., system potential energy and the configurational entropy, often feature in high VAMP-2 scoring crystallisation RCs provides validation that the characterisation of these processes, often adopted in simulations\cite{Sosso2016CrystalSimulations,Blow2021TheDetails,Wolde1997EnhancementFluctuations,Finney2022MultipleNucleation,Salvalaglio2015Molecular-dynamicsSolution,Zimmermann2015NucleationRates}, provide good CVs to reduce the high-dimension configuration space to a handful of relevant degrees of freedom and extract kinetic information.
\rev{In more complex systems, it may be necessary to incorporate additional CVs into the RC to describe how, for example, non-spherical monomers (perhaps with internal degrees of freedom), explicit solvent and impurities/additives affect nucleation. 
As there is no standard procedure to choose the best CVs to gain thermodynamic and kinetic information, trials of suitable functions to define (collective) molecular features must be performed.
CV accuracy can be affirmed using the analyses described in this work and elsewhere. \cite{Peters2016ReactionTests,Blow2021TheDetails}
Generally, the distribution of CV values representing the reactant, product and any intermediate states in a multi-step reaction pathway must be clearly distinguishable in CV space; hence, a multi-modal probability density of states should be apparent in the reaction coordinate. This is no guarantee, however, that the reaction coordinate is a good one to determine mechanisms and rates.}

The fact that the nucleation rates for the emergence of crystalline order in dense liquid intermediates are consistent, regardless of the number of CVs used to construct MSMs and determine timescales for these transitions, is a testament to the robustness of kinetic models constructed from CV combinations with a high VAMP-2 score.
A general conclusion from our analyses is that despite kinetic information being remarkably consistent in the MSMs constructed using TICA projections of $n=1-17$ CVs, quantitative thermodynamic information \emph{and} mechanistic insight are only accurately gained when a sufficiently large number of CVs are considered. 

From the majority of the MSMs constructed in this work, we were able to identify a crystallisation pathway progressing from the vapour phase to crystals via a dense liquid intermediate, with committor probabilities, spectral analysis and stationary distributions all indicating two bottlenecks to the formation of crystals: the condensation of the vapour to the liquid and rearrangement of particles in the liquid to form a crystal lattice, with the latter representing the rate-determining step.
Each of these two steps could, in principle, be described using their respective thermodynamic driving forces for nucleation, which form the basis of CNT. 
However, while a straightforward application of Gibbs' theory for nucleation might be possible to characterise the first step, the capillary approximation is likely to fail for crystal nucleation in the liquid, where we observed a population of nuclei with different defect densities and local crystalline arrangements. 

The model agreement across the range of dimensionalities ($n=1-17$) of CV spaces, and particularly the consistent prediction of nucleation rates, is a remarkable result that highlights the value of selecting combinations of crystallisation CVs that, for every $n$, maximise the VAMP-2 score.
We believe this approach is general and sufficiently transferable to support the study of other crystallisation or dissolution processes (where these events can be observed within reasonable simulation timescales) or to guide the choice of CVs used in enhanced sampling simulations of nucleation processes.

\begin{acknowledgments}
The authors acknowledge funding from an EPSRC Programme Grant (Grant EP/R018820/1) which funds the Crystallization in the Real World consortium. 
We thank members of the consortium for useful discussions.
The authors acknowledge the use of the UCL Myriad High Throughput Computing Facility (Myriad@UCL), and associated support services, in the completion of this work.
\end{acknowledgments}

\section*{Data Availability Statement}

LAMMPS and PLUMED input files, shell scripts used to automate data generation and interactive Python notebooks used in the analyses are available for download at https://github.com/aaronrfinney/VAMP-MSM.
PLUMED input files used in this work are also available via PLUMED-NEST (https://www.plumed-nest.org \cite{Bonomi2019PromotingSimulations}), the public repository for the PLUMED consortium, using the project ID: plumID:22.044.

\section{Supplementary material}
See the supplementary material for a detailed description of the collective variables (S1), the phase diagram of the colloidal system studied (S2), a summary of the highest scoring CV combinations (S3), and additional figures (S4). 

\section{\label{sec:references}References}
\bibliography{references}

\end{document}


\maketitle
{\begin{center}

\noindent \textit{Thomas Young Centre and Department of Chemical Engineering, University College London, London WC1E~7JE, United Kingdom}
\vspace{0.5 cm}

\noindent E-mail: a.finney@ucl.ac.uk; m.salvalaglio@ucl.ac.uk
\end{center}}
\tableofcontents

\clearpage
\section{Collective Variables}

The collective variables (CVs) evaluated in this work are listed in Table \ref{tab:table1}.
A minority of the CVs were computed during time integration by the LAMMPS package (v. 7Aug2019)\footnote{A. P. Thompson, H. M. Aktulga, R. Berger, D. S. Bolintineanu, W. M. Brown, P. S. Crozier, P. J. in 't Veld, A. Kohlmeyer, S. G. Moore, T. D. Nguyen, R. Shan, M. J. Stevens, J. Tranchida, C. Trott and S. J. Plimpton, ``LAMMPS - a flexible simulation tool for particle-based materials modeling at the atomic, meso, and continuum scale,'' Computer Physics Communications \textbf{271}, 10817 (2022).}; all other CVs were computed using PLUMED (v. 2.5)\footnote{G.A. Tribello, M. Bonomi, D. Branduardi, C. Camilloni, G. Bussi. ``PLUMED2: New feathers for an old bird,'' Computer Physics Communication \textbf{185}, 604 (2014).} by post-processing trajectories resulting from molecular dynamics.
The CV values were determined at every 100 steps during time integration.
See the ``Data Availability'' section of the main paper for instructions on how to access input files to compute the CVs.
In this section, we provide a very brief description of the functional forms for each CV and any algorithmic details needed to compute them.
Without exception, all quantities are reduced and, hence, are unitless, with conversion to physical units possible according to the \emph{lj} unit style in LAMMPS.
We direct the reader to the documentation associated with the LAMMPS and PLUMED software for a more complete description.

\begin{table}[h]
\center
\caption{\label{tab:table1}Collective variables (CVs) computed for crystallising trajectories in this work}
\begin{tabular}{lr}
\hline
\hline
CV&Label\\
\hline
\hline
Mean first-sphere coordination number & cn.mean \\
Number of particles in a condensed phase (CN > 3) & ncl \\
Number of particles in a solid-like phase (CN > 6) & ncs \\
Number of particles in the largest cluster & nclust1 \\
Mean Q4 Steindhardt bond order & Q4.mean \\
Mean local Q4 bond order & q4.mean \\
Number of coordinated particles with local Q4 < 0.3 & ncnq4 \\
Local average Q4 bond order & laQ4.mean \\
Mean Q6 Steindhardt bond order & Q6.mean \\
Mean local Q6 bond order & q6.mean \\
Number of coordinated particles with local Q6 > 0.7 & ncnq6 \\
Local average Q6 bond order & laQ6.mean \\
Pair entropy function & ent \\
System potential energy & ene\\
Number of particles not identified as fcc/hcp/bcc/ico \footnote{Evaluated using polyhedral template matching (PTM).} & non \\
Number of face-centred cubic particles $^\mathrm{a}$ & fcc \\
Number of hexagonal close-packed particles $^\mathrm{a}$ & hcp \\
Number of body-centred cubic particles $^\mathrm{a}$ & bcc \\
Number of icosahedral particles $^\mathrm{a}$ & ico \\
\hline
\hline
\end{tabular}
\end{table}

\subsection{CVs computed using LAMMPS}

\paragraph{ene}
The potential energy of the system must be calculated every step during time integration to determine the particle forces. 
As discussed in the ``Computational Details'' section of the main paper, a colloid/Yukawa potential was adopted to compute the potential energy between particles; this required a \emph{hybrid} pair style in LAMMPS and we used a distance cutoff of $12.5 \;\sigma$ to evaluate the particle energies and forces.
The colloid potential coefficients, labelled to be consistent with the LAMMPS documentation, were A = 53, $\sigma$ = 1, d1 = 5, d2 = 5 and cutoff = 12.5.
In addition, the Yukawa potential coefficients were A = 20 and cutoff = 12.5.

\paragraph{non, fcc, hcp, bcc, ico}
These CVs are computed using the polyhedral template matching  (PTM) algorithm implemented in LAMMPS, which adopts the method of Larsen et al.\footnote{ P. M. Larsen, S. Schmidt and J. Schiøtz, ``Robust structural identification via polyhedral template matching,'' 
Modelling and Simulation in Materials Science and Engineering \textbf{24}, 055007 (2016).
 }
This method involves computing the root mean squared deviation, RMSD, between the local particle structural environment and reference face centered cubic, hexagonal close packed, body centered cubic and icosahedral crystal lattices according to,
\begin{equation}
    \mathrm{RMSD}(\mathbf{r},\mathbf{v}) = \mathrm{min} \left( \frac{1}{N} \sum_{i=1}^N || s [\vv{r}_i - \bar{\mathbf{r}}] - \mathbf{Q} \cdot \vv{v}_i ||^2
    \right)^\frac{1}{2}
\end{equation}
In the equation, $N$ is the number of particles, while $\mathbf{r}$ and $\mathbf{v}$ are vectors of the particle and crystal lattice Cartesian coordinates, respectively. 
LAMMPS optimises the scaling factor, $s$, and rotation matrix, $\mathbf{Q}$, to evaluate the RMSD between the particle coordinates, $\vv{r}$, and the reference lattice, $\vv{v}$.
A match to the most probable crystal structure is found for the lattice with lowest RMSD.
An RMSD threshold of 0.22 was used to identify particles with a local structure that did not match any of the reference crystal structures i.e., non.
This was chosen by constructing colloid crystal lattices in LAMMPS with lattice constants equal to the colloid particle-particle interaction distances in the crystals that emerged from simulations, ensuring that the PTM algorithm produced the correct number of particles for the chosen crystal structure.
An additional step was required to count the number of particles identified as non, fcc, hcp, bcc or ico for each frame in the resulting output file.

\subsection{CVs computed using PLUMED}

\paragraph{cn.mean}
The mean particle first-sphere coordination number (CN) was computed as,
\begin{equation}
    \overline{\mathrm{CN}}=\frac{1}{N}\sum_i^N \sum_{j \neq i}^N \; \frac{f(r_{ij})-f(r_{max})}{1-f(r_{max})}
    \label{eq:cn}
\end{equation}
where $r_{ij}$ is the distance between colloids $i$ and $j$, and $r_{max}$ was 7.
This makes use of the continuous \emph{rational} switching function implemented in PLUMED to identify particles within a coordination sphere,
\begin{equation}
    f(r)=\frac{1-(\frac{r}{r_0})^p}{1-(\frac{r}{r_0})^{q}}
    \label{eq:switch}
\end{equation}
where $r_0=6.4 \sigma$ and 1 for $f(r_{ij})$ and $f(r_{max})$, respectively, $p=6$ and $q=12$.
$r_0=6.4 \sigma$ was chosen as this distance marks the first minimum in particle-particle radial distribution functions computed from the end of crystallising trajectories.
Use of $f(r_{max})$ in Equation \ref{eq:cn} results in shifting and stretching of $f(r_{ij})$ to truncate the function for first sphere colloid-colloid coordination at a reduced radial distance of $7 \;\sigma$. 

\paragraph{ncl, ncs}
To compute ncl and ncs, the coordination number (CN) for all particles was evaluated using a formula similar to Equation \ref{eq:cn}, but here only a single sum over all particles is computed; hence, a CN distribution for all of the particles is obtained.
We then applied a filter, with form $1-f(r=x)$, to identify those particles with a CN greater than $x=3$ or $x=7$, the total numbers of which provided ncl and ncs, respectively.

\paragraph{nclust1}
As well as counting particles based on some property of the local structure, it is helpful to identify all particles which are connected in the largest cluster.
Here, we made use of the adjacency matrix module in PLUMED.\footnote{G. A. Tribello, F. Giberti, G. C. Sosso, M. Salvalaglio and M. Parrinello, ``Analyzing and Driving Cluster Formation in Atomistic Simulations,'' Journal of Chemical Theory and Computation, \textbf{13}, 1317–1327 (2017).}
First, a contact matrix is constructed for all particles connected within their first coordination sphere according to Equation \ref{eq:switch}.
A depth first search graph reduction algorithm\footnote{S. Even, ``Grap Algorithms 2nd Ed.,'' Cambridge University Press (2011).} then identifies all sets of connected particles.
The number of particles in each set is computed to identify the largest cluster.

\paragraph{Q4.mean, Q6.mean}
In order to quantify symmetries in the arrangement of particles in their first coordination spheres and, therefore, determine local order, CVs based on Steinhardt bond orientational order parameters, $Q$, were used.\footnote{P. J. Steinhardt, D. R. Nelson, and M. Ronchetti, ``Bond-orientational order in liquids and glasses,'' Physical Review B \textbf{28}, 784 (1983).}
Computing $Q$ to the $l$-th order involves evaluating spherical harmonic functions, $Y_{lm}(\vv{r}_{ij})$, from the Cartesian coordinates of particle $i$ and its first sphere particle connections (identified using Equation \ref{eq:switch}):
\begin{equation}
    y_{lm}^i = \frac{\sum_j^N f(r_{ij}) Y_{lm}(\vv{r}_{ij})}{\sum_j^N f(r_{ij})}
\end{equation}
In our case, $l=4$ and 6 and when computing the complex vectors $y_{lm}$; hence, there are 9 ($m=\{-4, \dots{}, 0, \dots{}, +4\}$) or 13 ($m=\{-6, \dots{}, 0, \dots{}, +6\}$) spherical harmonic functions, and the mean $Q_{l}$ is evaluated as,
\begin{equation}
    \overline{Q_{l}} = \frac{1}{N} \sum_i^N \left( \sum_{-m}^m y_{lm}^{i*} \;  y_{lm}^{i} \right)^\frac{1}{2}
\end{equation}

\paragraph{q4.mean, q6.mean}
The mean $Q_l$ quantifies the global order in the system for all atoms and is indicative of the first-sphere coordination symmetry in the dominant phase. 
This CV is, therefore, rather insensitive to the emergence of order which arises in a small region of the simulation cell during crystallisation.
The mean local bond orientational order parameter, $\overline{q}_l$, instead quantifies how order in the particles coordinated to a central particle $i$ is replicated.
This is achieved by computing,
\begin{equation}
    \overline{q_{l}} = \frac{1}{N} \sum_i^N \left( \frac{\sum_j f(r_{ij}) \sum_{-m}^m y_{lm}^{i*} \;  y_{lm}^{j}}{\sum_j f(r_{ij})}
    \right)
\end{equation}

\begin{figure}
\includegraphics[width=\textwidth]{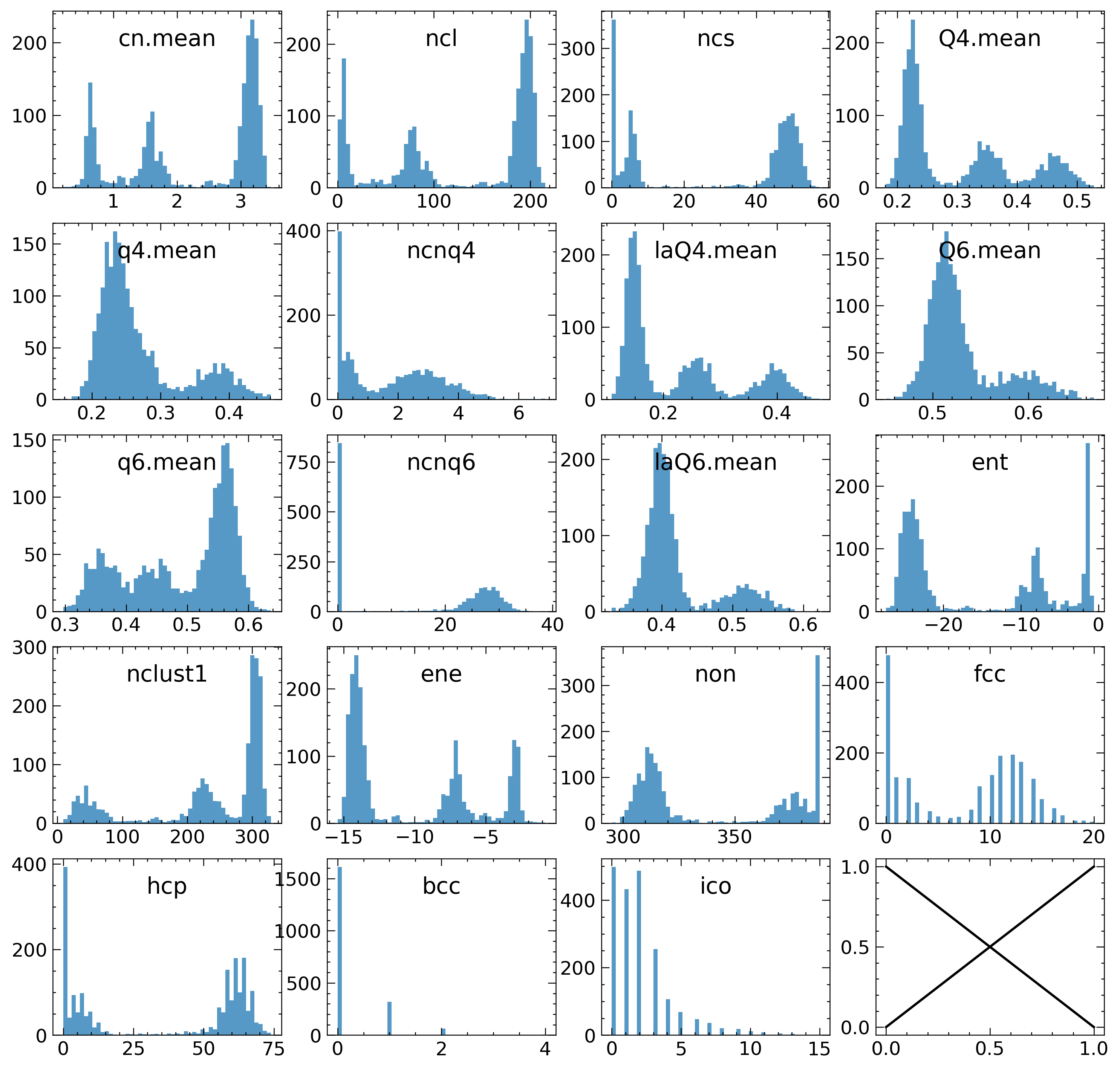}
\caption{\label{fig:si-cv-hist} Histograms for the CVs computed in this work for a single crystallising trajectory (for clarity) with the sub-headings indicating the CVs.}
\end{figure}

\paragraph{ncnq4,ncnq6}
In classical nucleation theory, the size of the emerging phase defines the reaction coordinate.
It is possible to count the total number of particles with a level of order beyond some threshold by making use of ${q}_{l}$ particle distributions.
Here, a filtering function of the form, 
\begin{equation}
    s(r) =  \exp \left( \frac{-(r-d_0)^2}{2r_0^2}
    \right)
\end{equation}
is used to identify particles with ${q}_{l} < 0.3 \; (r_0 = 0.3)$ and ${q}_{l} > 0.7 \; (r_0 = 0.7)$.
In the first case, we apply the filter using $s(r){q}_{l}$, while in the latter we use $(1-s(r)){q}_{l}$. 
The values of 0.3 and 0.7 were informed by the ${q}_{l}$ distributions (see Figure \ref{fig:si-cv-hist}) and confirmed by embedding crystals into vapour phases to check that the CV correctly identified particles in the crystallites.
It is important to recognise that the correct identification of crystal particles at the interface between two phases is very sensitive to the choice of these thresholds.

After selecting those particles with crystalline order, we compute the first-sphere coordination distribution for these particles using a function similar to the one in Equation \ref{eq:cn} (with $r_0=6.4$ and $r_{max}=7$), but here we only compute the single sum over $j$ particles.
As in the case of computing the CN between all particles, we chose $r_0=6.4 \sigma$ and $r_{max}=7 \sigma$ and we apply an additional \emph{rational} filter (i.e., Equation \ref{eq:switch}) where $f(r=x)$, with $x=6$.
This allowed us to identify particles with crystalline order and with a CN > 6; hence, we only select those particles which are within a crystal nucleus.
In doing so, we aim to differentiate crystal nuclei from single particles and very small particle clusters which may rapidly fluctuate between amorphous and crystal-like local structures.
We can then sum the number of particles to estimate the size of the crystal nucleus.
In more complex systems or at higher supersaturations, an additional clustering step may be needed to identify different nuclei and determine their sizes; however, observations of the trajectories and the mechanism for crystallisation in this system confirmed that this step was not necessary.

\paragraph{laQ4.mean, laQ6.mean}
Given the potential for $Q_l$ CVs to fail to identify the onset of crystallisation, we also computed the mean local average $Q_l$ which was shown to better differentiate between liquid and solid states in Lennard-Jones systems.\footnote{W. Lechner and C. Dellago ``Accurate determination of crystal structures based on averaged local bond order parameters,'' The Journal of Chemical Physics \textbf{129}, 11 (2008).} 
This method of computing local average CVs is computationally cheaper than evaluating $q_l$, as it simply takes the distributions of $Q_l$ as inputs to examine the local particle structre:
\begin{equation}
    \overline{Q_{l(r_{ij}<r_{max})}} = \frac{1}{N}  \sum_i^N \left(\frac{Q_l^i + \sum_j^N f(r_{ij}) Q_l^j}{1+ \sum_j^N f(r_{ij})} \right)
\end{equation}
Again we only consider the first coordination sphere, therefore $r_0=6.4 \sigma$ and $r_{max}=7 \sigma$ in Equation \ref{eq:switch}.

\paragraph{ent}
The pair entropy function, $S_2$, is a CV that was used to enhance the  sampling of crystallisation events in liquid metals.\footnote{P. M. Piaggi, O. Valsson, and M. Parrinello, ``Enhancing Entropy and Enthalpy Fluctuations to Drive Crystallization in Atomistic Simulations'' Physical Review Letters 119, 015701 (2017)}
It makes use of an approximate function for the entropy of the system based on a pair correlation function, $g(r)$:
\begin{equation}
    g(r) = \frac{1}{4 \pi N \rho^* r^2} \sum_i^N \sum_{j \neq i}^N \exp \left( \frac{-(r-r_{ij})^2}{2 \xi ^2}
    \right)
\end{equation}
where $\rho^*$ is the system number density, $k_\mathrm{B}$ is Boltzmann's constant and $\xi = 0.01$.
A limit of $r_{lim}<12 \sigma$ was imposed to reduce the cost of CV computations.
The approximate entropy is then given by,
\begin{equation}
    S_2 = -2 \pi \rho k_\mathrm{B} \int_0^{r_{lim}} r^2 [g(r) \ln g(r) - g(r) +1] \;dr
\end{equation}
The assumption that the entropy accounts only for those states associated with two-body correlations fails for most multi-component systems, but for the simple, spherical particle system adopted here, it is reasonable.

\begin{figure}
\includegraphics[width=\textwidth]{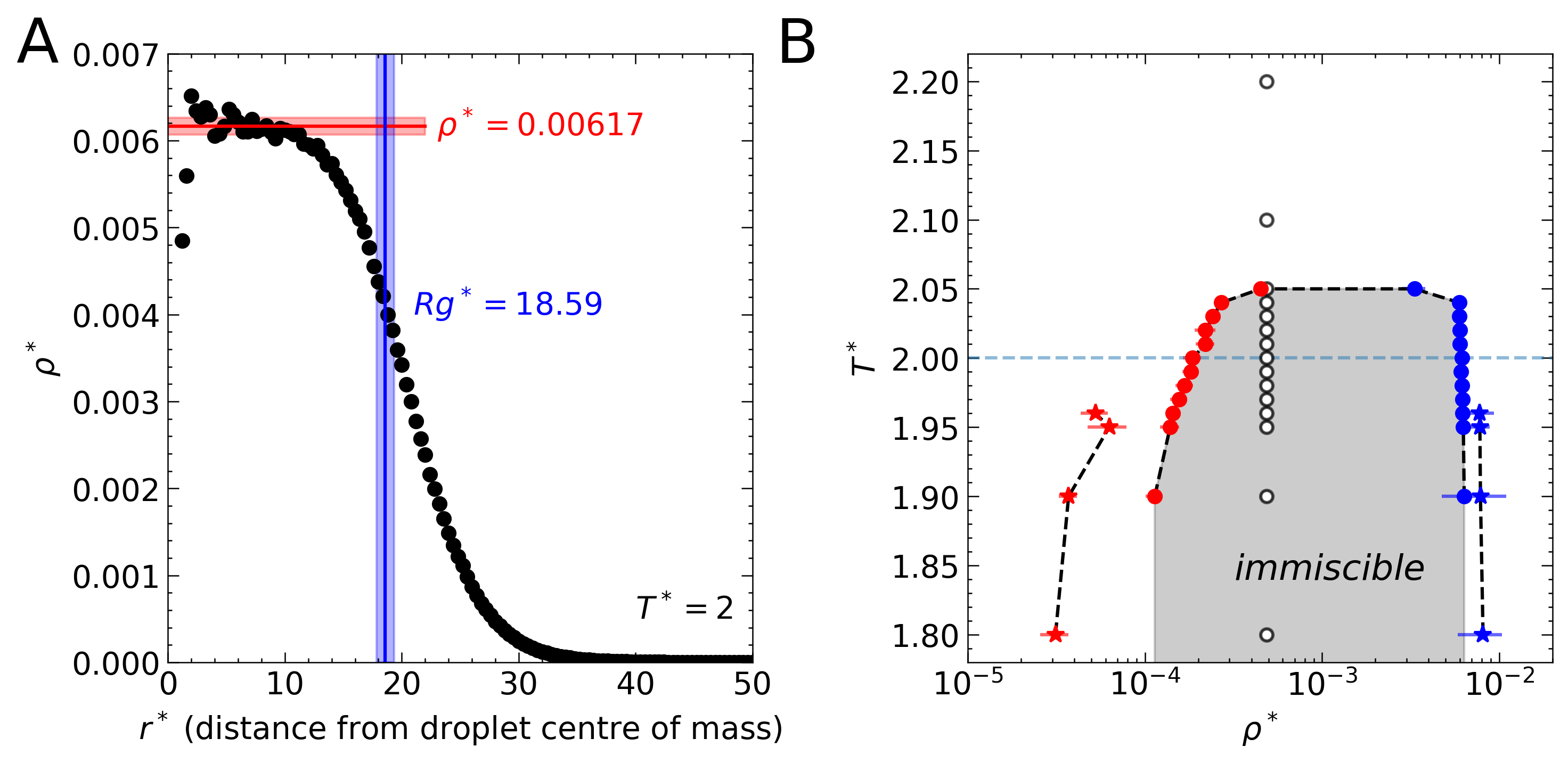}
\caption{\label{fig:phase-diag} \rev{A) The radial density profile for a dense liquid droplet observed in a steady state with a vapour phase when $T^* = 2$. The mean reduced radial density for the liquid is shown by the red line, while the mean radius for the droplet, as determined by its radius of gyration ($R_g$), is provided by the blue line. Shaded regions indicate uncertainties associated with one standard deviation in the mean data. B) The $T^*-\rho^*$ phase diagram estimated from simulations initiated at the $T^*$ and $\rho^*$ marked by the black circles. Where phase separation produced a dense liquid droplet in pseudo-equilibrium with a vapour phase, we highlight the resulting $T^*$ and $\rho^*$ using blue and red circles for the condensed phase and vapour phase branches of the immiscible region, respectively. We also include the $T^*-\rho^*$ crystal-vapour equilibrium densities, represented by stars. Error bars indicate one standard deviation in the mean reduced densities. Black dashed lines provide an approximate guide for the binodal, while the blue dashed line highlights the $T^*$ adopted in the simulations discussed in the main paper. The shaded area highlights the immiscible region where vapour-liquid phase separation is likely to occur.}}
\end{figure}

\rev{
\clearpage
\section{Colloid Phase Diagram}
To estimate the phase behaviour for the model system adopted, we performed an additional 15 simulations. All of these were prepared using the same starting configuration by randomly placing particles on a sparse face centred cubic lattice with a reduced particle density, $\rho^* = 0.000485$, i.e., matching the simulation preparation described in the main paper. Here, however, we varied the temperature in simulations where $T^* = 1.8$, 1.9, 1.95, 1.96, 1.97, 1.98, 1.99, 2, 2.01, 2.02, 2.03, 2.04, 2.05, 2.1 and 2.2. Analysis of the resulting trajectories indicated the formation of a relatively long-lived dense liquid droplet when $T^* = 1.9-2.05$. Below $T^* = 1.9$, the system spontaneously separated into crystal and vapour phases. On the other hand, above $T^* = 2.05$, no phase separation was evident on the timescale of the simulations; instead, a high-density vapour phase---containing spatially extended particle clusters with rapidly fluctuating size and structure---was apparent. Finally, at $T^* = 1.8-1.96$, crystallisation occurred in the dense liquid droplets. We can, therefore, identify the two-step crystallisation pathway in three of the trajectories when $T^* = 1.9$, 1.95 and 1.96.  

We determined the reduced density of the dense liquid droplets and nanocrystals by first computing the average particle number density as a function of the distance from the centre of mass of the largest particle clusters in trajectory windows where these phases occur (the number of frames used here was informed by plateau regions in average particle potential energy time series data).  An example radial density profile for a droplet which emerged when $T^* = 2$ is provided in Figure \ref{fig:phase-diag} A. The mean $\rho^*$ was then computed from the core region of the droplets/crystals (see the red line in Figure \ref{fig:phase-diag} A). The density for the corresponding vapour phase was determined as $(N-n)/(V-Rg^3)$, where $N$ is the total number of particles, $n$ is the mean number of particles in the condensed phase, $V$ is the total volume of the simulation cell and $Rg$ is the mean radius of gyration of the droplet or crystal. $Rg$ approximately marks the radial distance where the density profile reduces to half its maximum (see Figure \ref{fig:phase-diag} A), and so provides a good estimate for the condensed phase radius. 

From the densities of the vapour phase, liquid droplets and crystals, we constructed the $T^*- \rho^*$ phase diagram provided in Figure \ref{fig:phase-diag} B. Interpolating between the points marking the dense liquid and vapour phase at steady state (circles) provides an approximate boundary for the immiscible region where fluid-fluid demixing occurs, known as the binodal. The shape of the binodal suggests an upper critical temperature ($T_c^*$) around $T^*= 2.05$ when $\rho^* = 10^{-3}$. Taking $T_c^*=2.05$, this suggests that the crystallisation simulations discussed in the main paper are performed at $T^*/T_c^* = 0.976$. Such close proximity to the critical point makes liquid-like intermediates more probable in the crystallisation pathway, according to earlier simulation studies.\footnote{P. R. ten Wolde and D. Frenkel, ``Enhancement of Protein Crystal Nucleation by Critical Density Fluctuations'', Science \textbf{277}, 1975–1978 (1997).
}

}
\clearpage
\section{Highest Scoring CVs}
\begin{table}[h]
\center
\caption{\label{tab:table1}The RCs for each dimension, $n$, with the highest VAMP-2 score.}
\begin{tabular}{p{0.1\linewidth}p{0.9\linewidth}}
\hline
\hline
$n$ & CVs\\
\hline
\hline
1 & ncs \\
2 & ncs, ncnq6 \\
3 & ncs, ncnq6, nclust1 \\
4 & ncs, ncnq6, nclust1, fcc \\
5 & ncs, ncnq6, ent, ene, fcc \\
6 & ncl, ncs, ncnq6, ent, nclust1, fcc \\
7 & ncl, ncs, q4.mean, ncnq6, ent, nclust1, fcc \\
8 & ncl, ncs, q4.mean, laQ4.mean, ncnq6, ent, nclust1, fcc \\
9 & ncl, ncs, q4.mean, laQ4.mean, ncnq6, ent, nclust1, ene, fcc \\
10 & ncl, ncs, q4.mean, laQ4.mean, ncnq6, ent, nclust1, ene, fcc, hcp \\
11 & ncl, ncs, q4.mean, ncnq4, laQ4.mean, ncnq6, ent, nclust1, ene, fcc, hcp \\
12 & ncl, ncs, q4.mean, ncnq4, laQ4.mean, ncnq6, laQ6.mean, ent, nclust1, ene, fcc, hcp \\
13 & ncl, ncs, q4.mean, ncnq4, laQ4.mean, Q6.mean, ncnq6, laQ6.mean, ent, nclust1, ene, fcc, hcp \\
14 & ncl, ncs, Q4.mean, q4.mean, ncnq4, laQ4.mean, Q6.mean, ncnq6, laQ6.mean, ent, nclust1, ene, fcc, hcp \\
15 & ncl, ncs, Q4.mean, q4.mean, ncnq4, laQ4.mean, Q6.mean, q6.mean, ncnq6, laQ6.mean, ent, nclust1, ene, fcc, hcp \\
16 & cn.mean, ncl, ncs, Q4.mean, q4.mean, ncnq4, laQ4.mean, Q6.mean, q6.mean, ncnq6, laQ6.mean, ent, nclust1, ene, fcc, hcp \\
17 & cn.mean, ncl, ncs, Q4.mean, q4.mean, ncnq4, laQ4.mean, Q6.mean, q6.mean, ncnq6, laQ6.mean, ent, nclust1, ene, fcc, hcp, non \\
\hline
\hline
\end{tabular}
\end{table}

\clearpage
\section{Additional Figures}
\clearpage

\begin{figure}
\includegraphics[width=\textwidth]{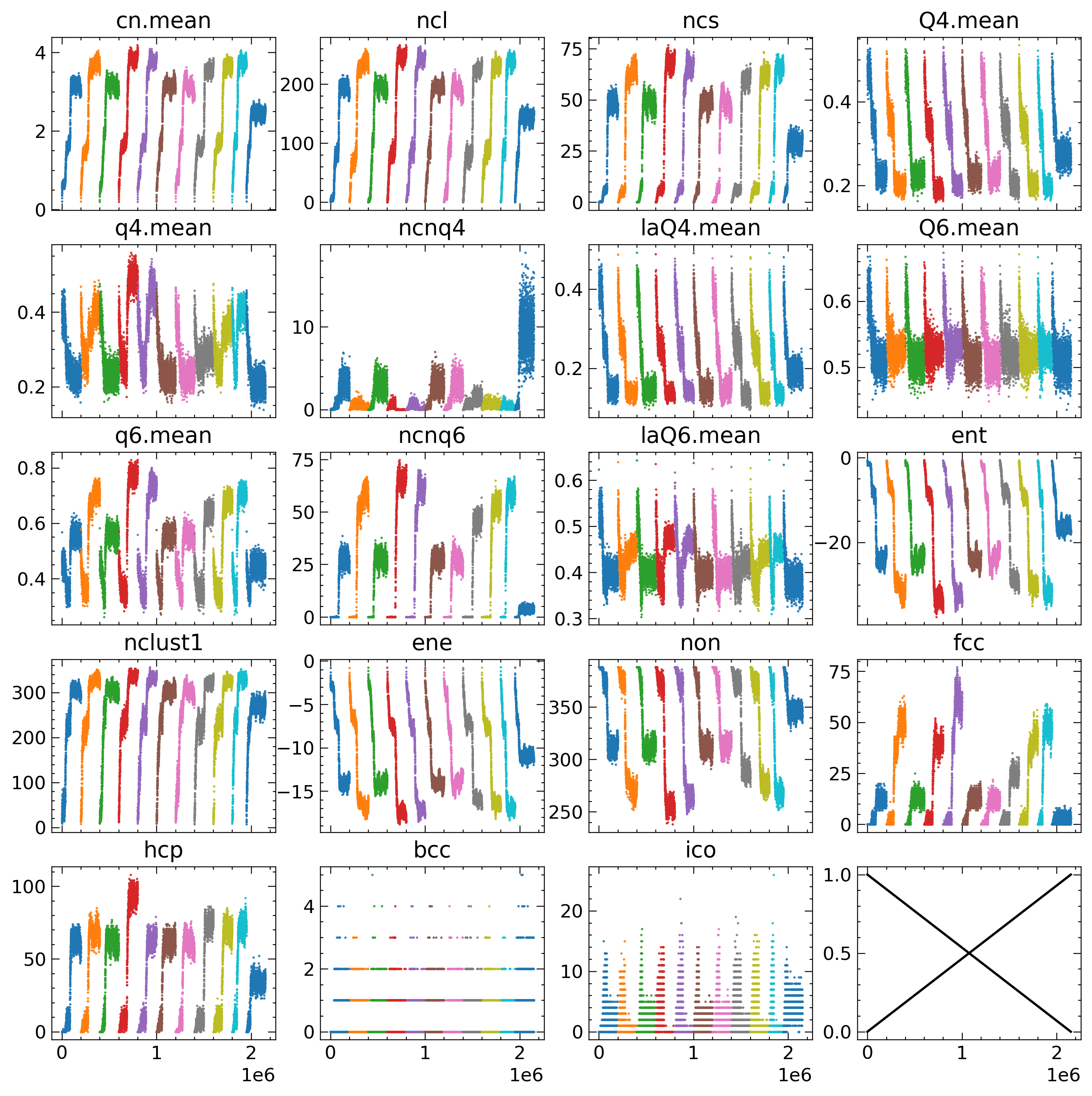}
\caption{\label{fig:si-cv-time} Time series data for the CVs computed in this work indicated by the sub-headings. Different crystallising trajectories are indicated by the different colours. Trajectories are concatenated, such that the step number on the $x$-axis runs continuously over trajectories and represents the cumulative frame number.}
\end{figure}

\begin{figure}
\includegraphics[width=\textwidth]{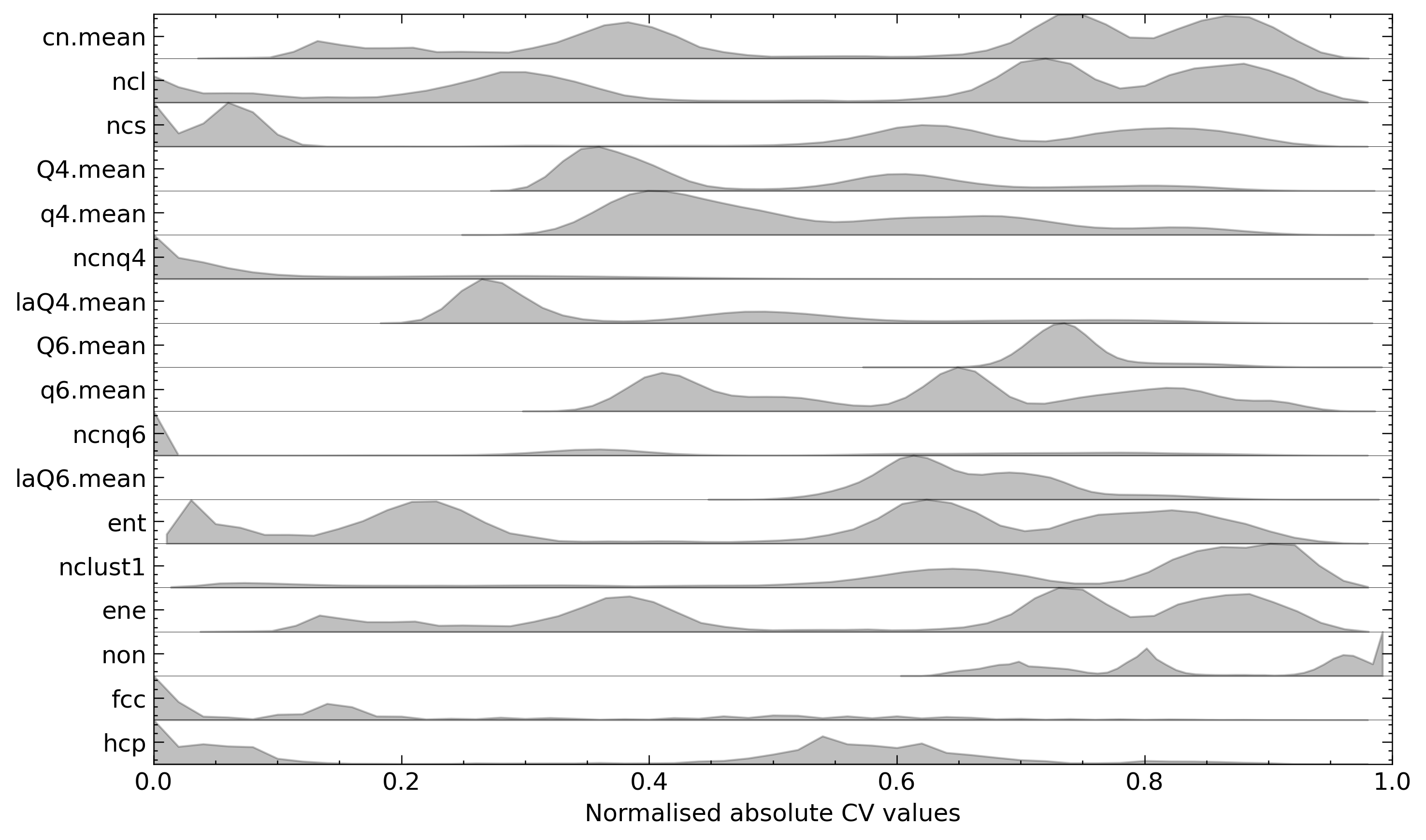}
\caption{\label{fig:si-cv-dist-scaled} Distributions for the scaled, absolute CV coordinates used to compute VAMP-2 scores.}
\end{figure}

\begin{figure}
\includegraphics[width=\textwidth]{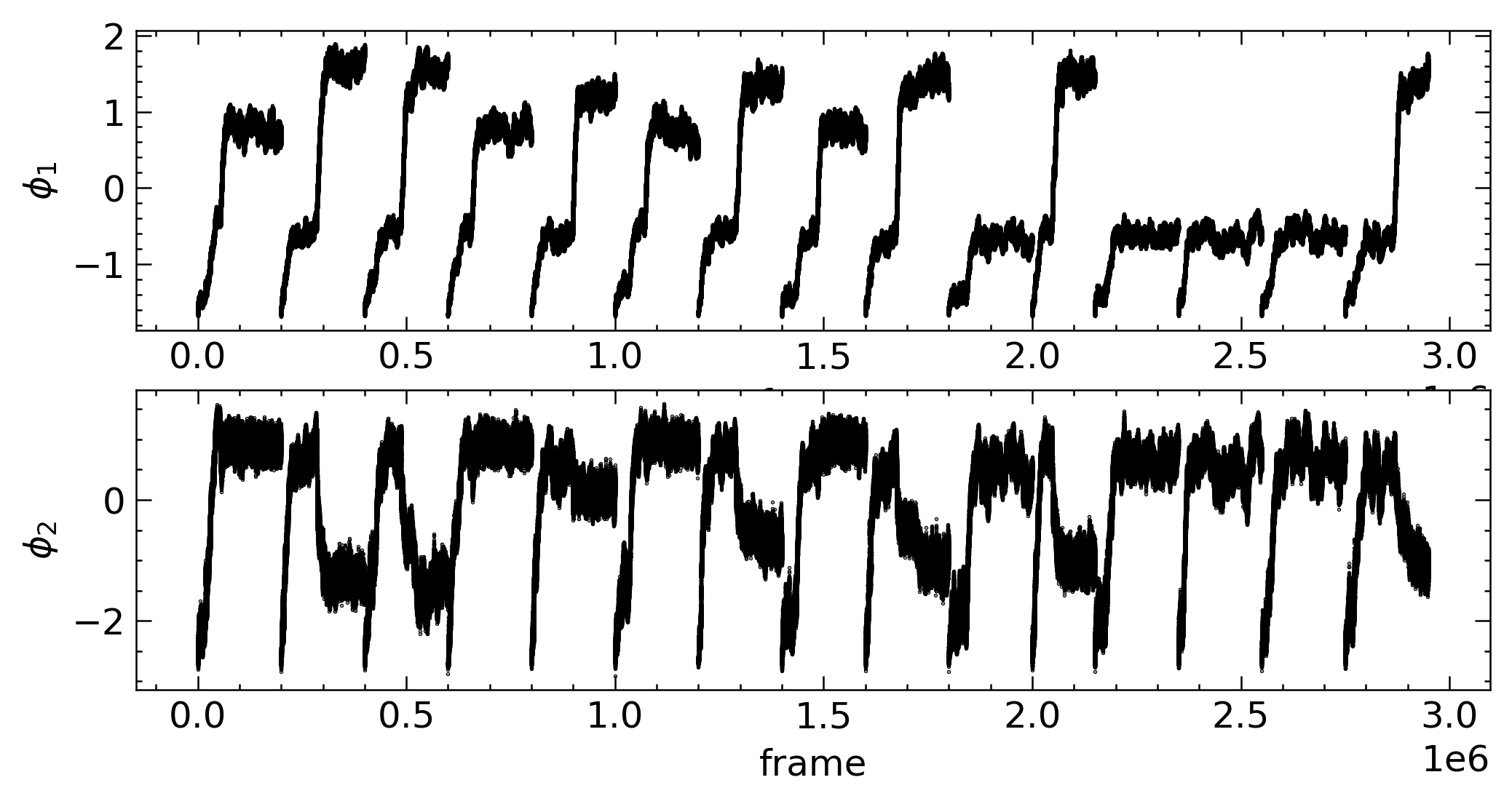}
\caption{\label{fig:si-tics-traj} Concatenated simulation trajectories projected onto 17 CVs and reduced to the two TICA dimensions, $\phi_1$ and $\phi_2$.}
\end{figure}

\begin{figure}
\centering
\includegraphics[width=0.5\textwidth]{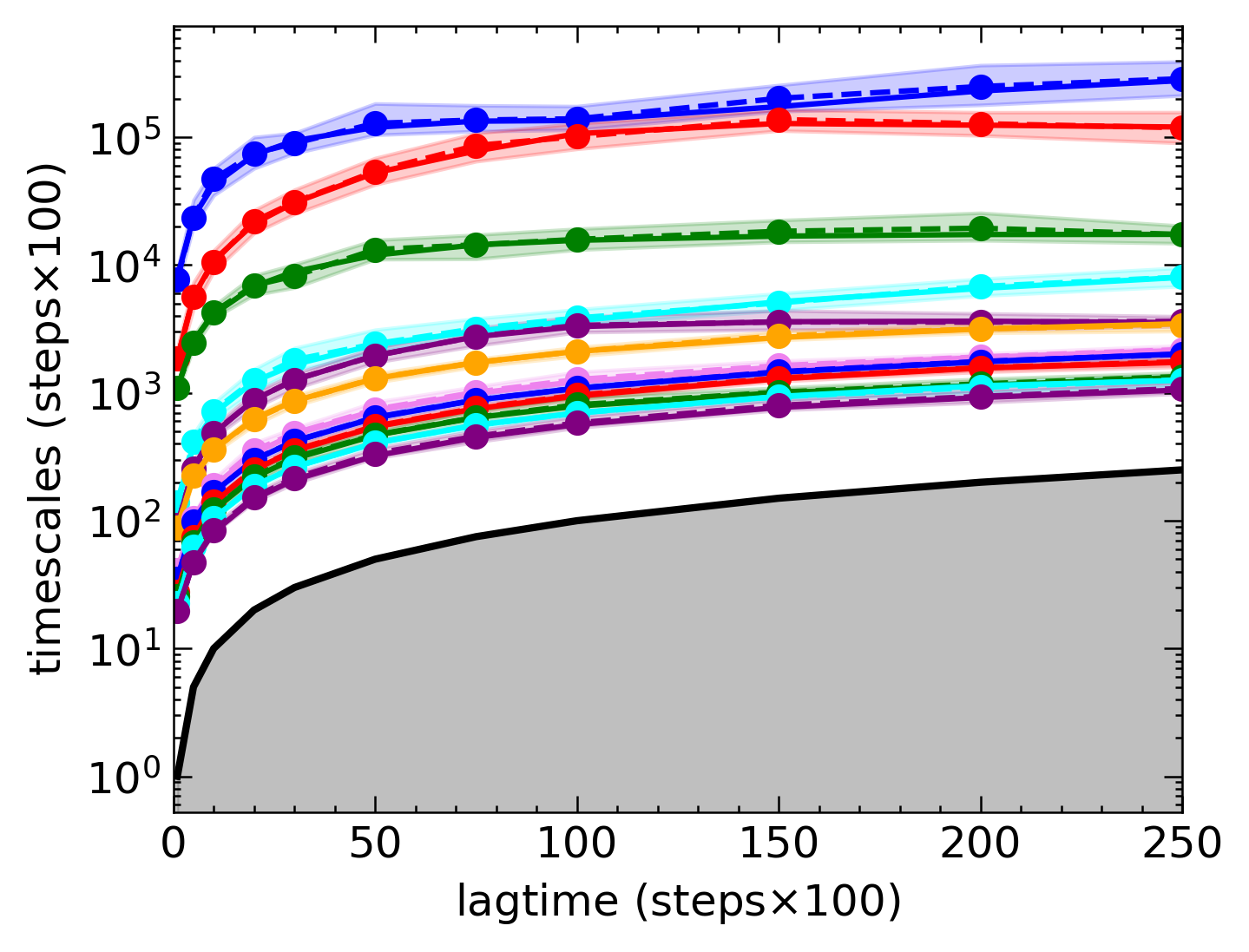}
\caption{\label{fig:si-BayesMSM-its} Implied timescales for the twelve slowest dynamical modes as a function of lagtime used to build Bayesian MSMs using a 2D TICA decomposition of $n=17$ CVs. Shaded regions indictate a 95\% condfidence interval.}
\end{figure}

\begin{figure}
\centering
\includegraphics[width=\textwidth]{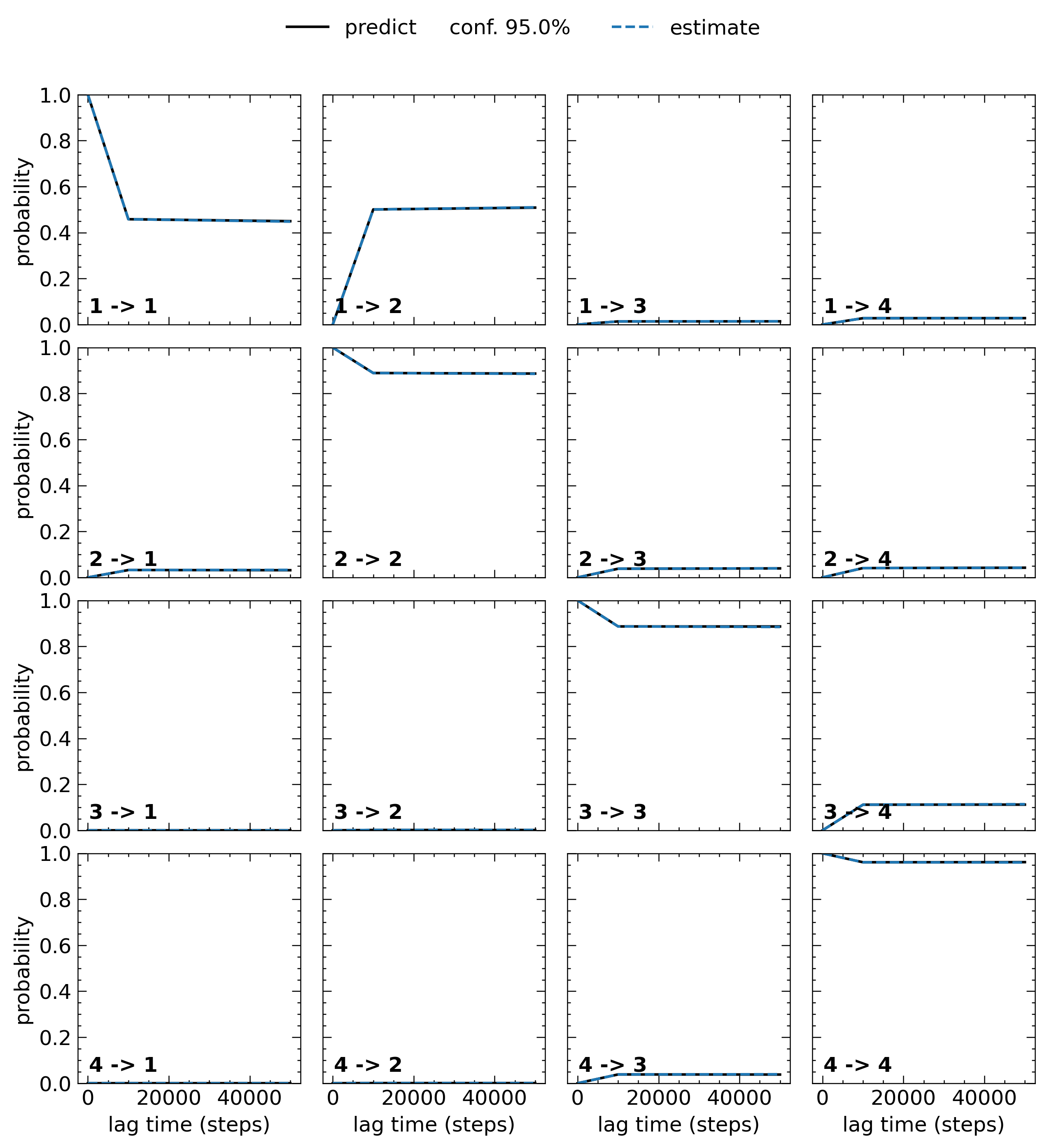}
\caption{\label{fig:si-CK} Results from a Chapman-Komogorov test applied to the Bayesian MSM constructed using a 2D TICA decomposition of the $n=17$ CVs. Model predictions and estimates are provided, and uncertainties from a 95\% confidence interval in the model predictions are not visible on this scale. }
\end{figure}

\begin{figure}
\centering
\includegraphics[width=\textwidth]{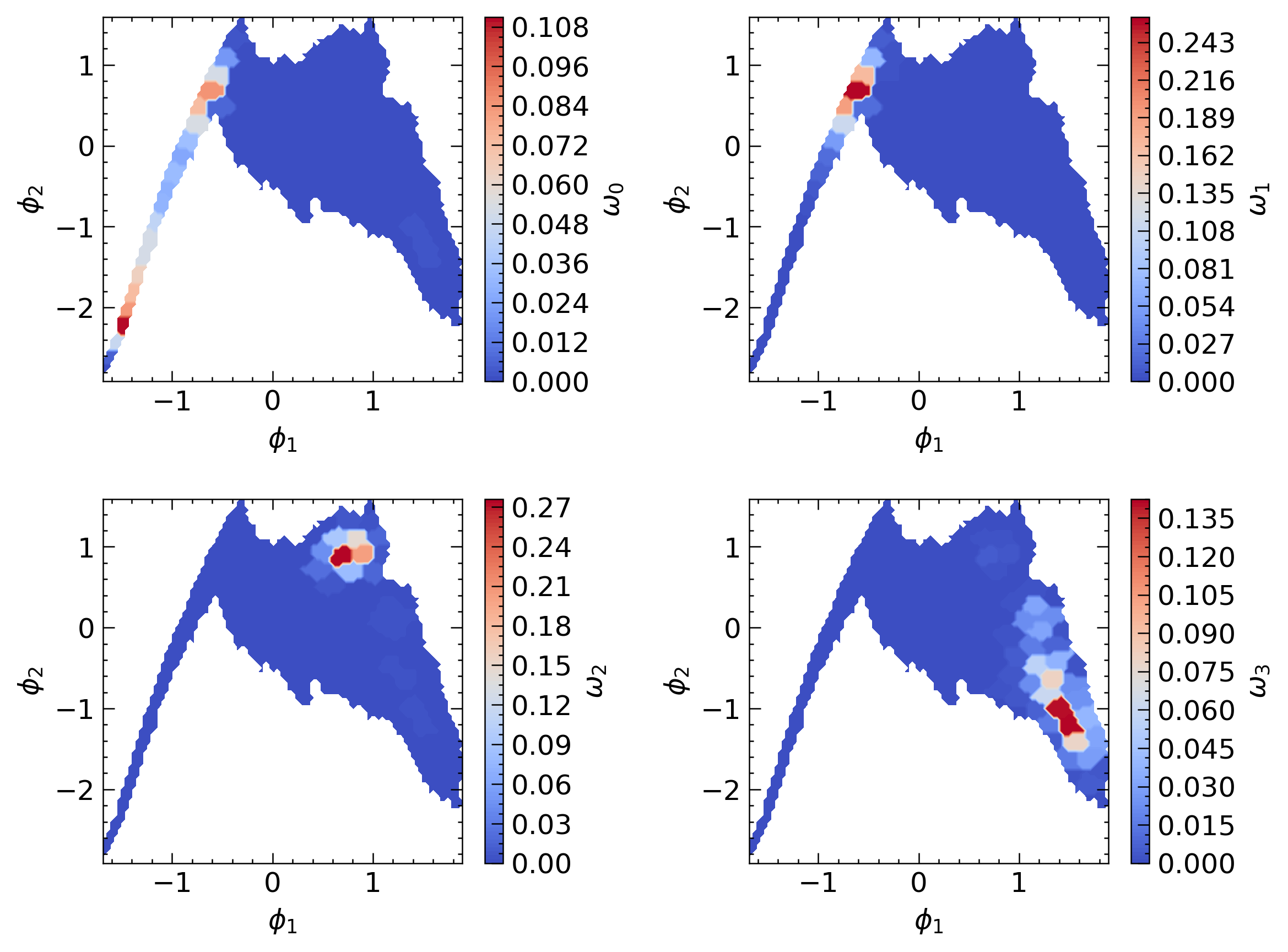}
\caption{\label{fig:si-metastable-states} PCCA+ assignment of partitions to four (meta)stable states, $\omega$, in a MSM constructed using 2D TICA decomposition of $n=17$ CVs. }
\end{figure}

\begin{figure}
\centering
\includegraphics[width=\textwidth]{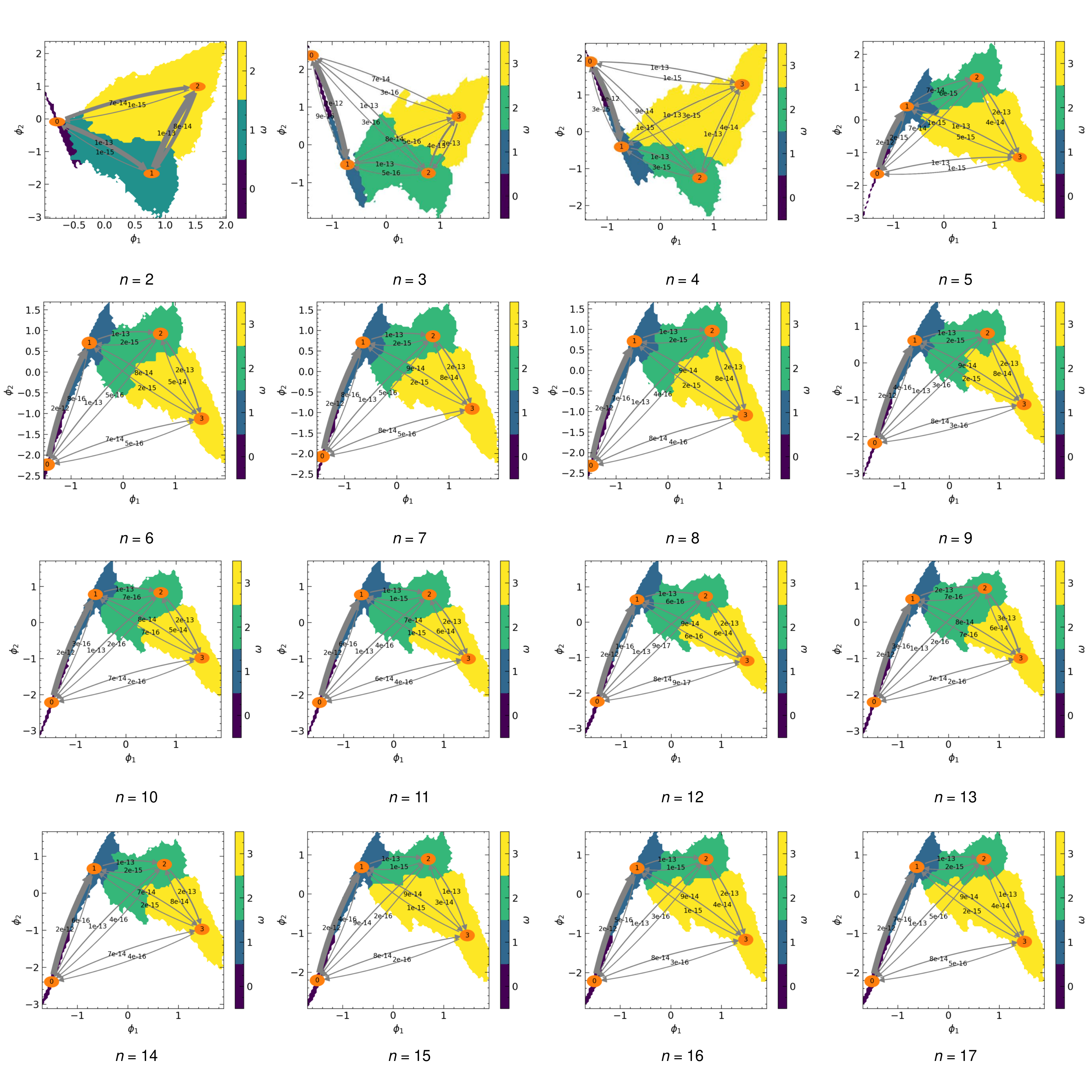}
\caption{\label{fig:si-tica-pcca} Metastable states $\omega$ and rates (in $\sigma^{-3} \; \mathrm{steps}^{-1}$ units) for transitions between them, indicated by the arrow labels, for MSMs using 2D TICA decompositions of the highest $R_2$ scoring RCs comprising $n=2-17$ CVs. }
\end{figure}

\begin{figure}
\centering
\includegraphics[width=\textwidth]{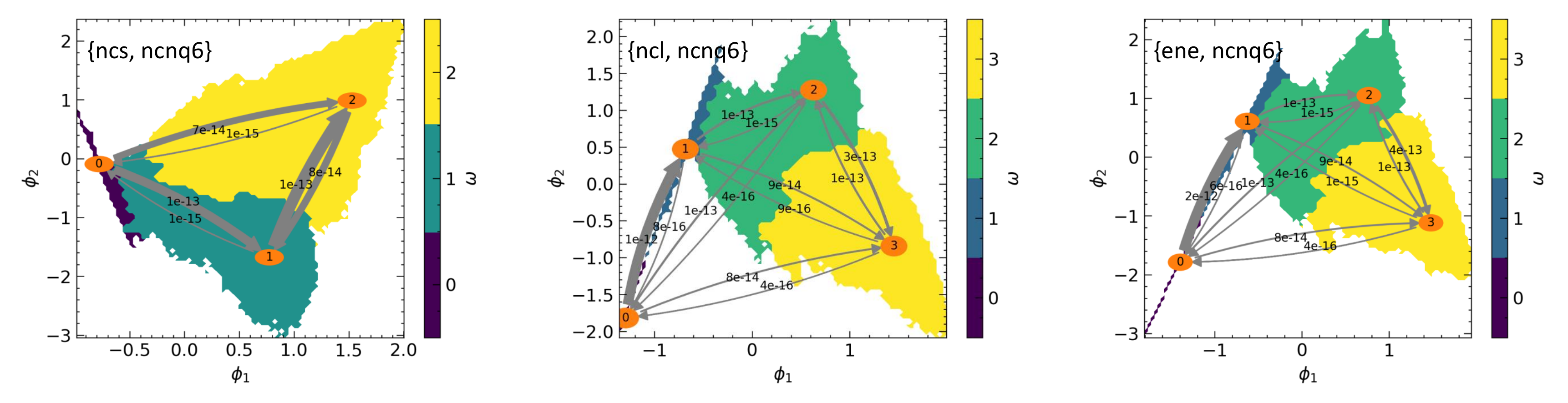}
\caption{\label{fig:si-tica-pcca} Metastable states $\omega$ and rates (in $\sigma^{-3} \; \mathrm{steps}^{-1}$ units) for transitions between them, indicated by the arrow labels, for MSMs using 2D TICA decompositions of the highest $R_2$ scoring RCs comprising $n=2$ CVs, indicated by the labels inset. }
\end{figure}

\begin{figure}
\centering
\includegraphics[width=\textwidth]{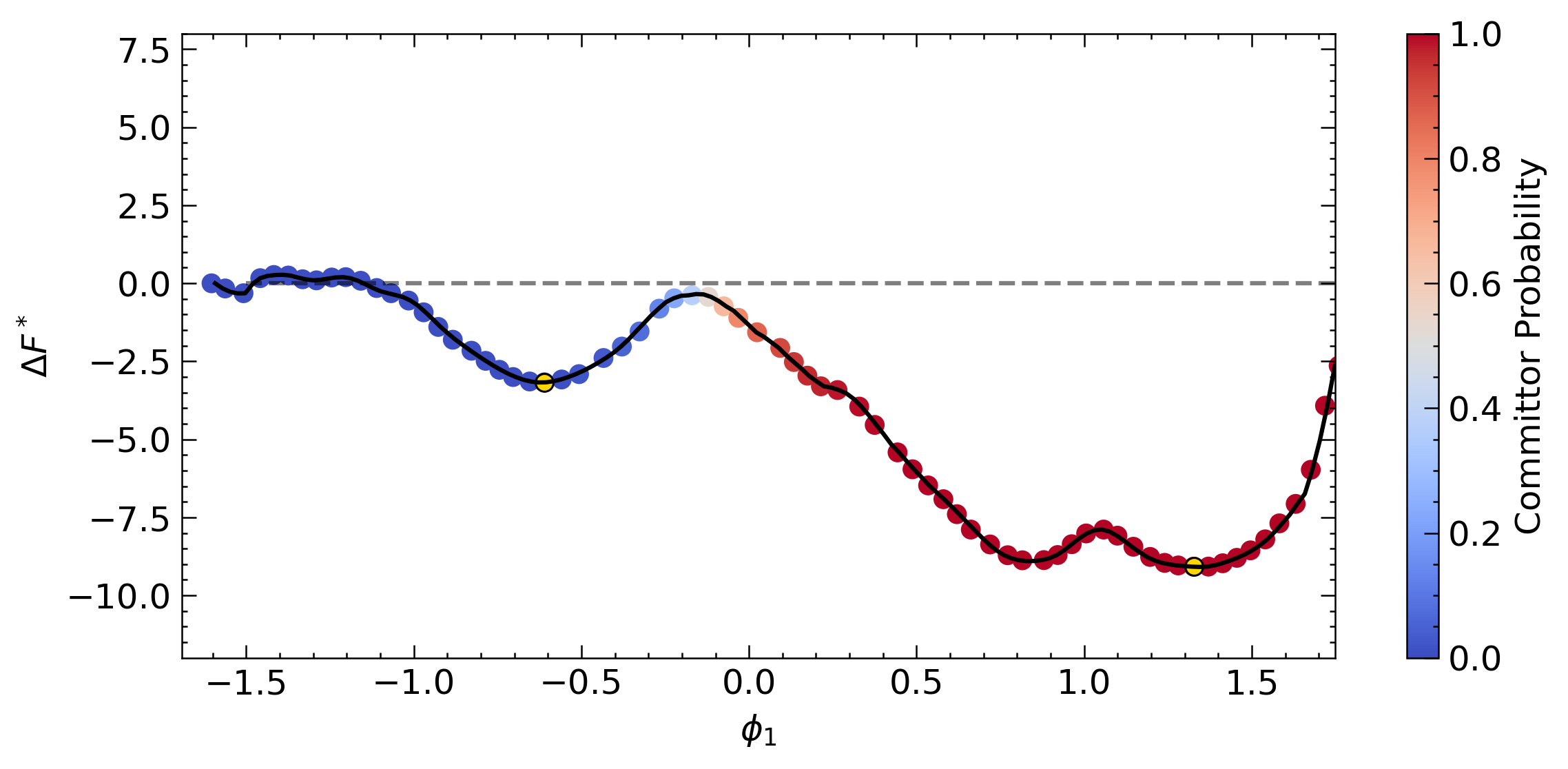}
\caption{\label{fig:si-1D} Relative free energy ($\Delta F^*$) as a function of $\phi_1$, the TICA coordinate onto which the ncl CV time series data are projected. The circles indicate the cluster centres, coloured according to their probability to commit to the yellow cluster centre in the C$_3$ minimum from the DLD minimum.  }
\end{figure}